\documentclass[twocolumn,longabstract,numberedappendix,twocolappendix]{aastex631}
\usepackage{amsmath,amssymb,commath}
\usepackage[displaymath,mathlines]{lineno}
\usepackage{tikz}
\usetikzlibrary{matrix,positioning,decorations.pathreplacing}
\usepackage{afterpage}
\usepackage{rotating}
\usepackage{graphicx}
\usepackage{color}
\usepackage[normalem]{ulem}
\usepackage{subfloat}
\usepackage{verbatim}
\usepackage{makecell}
\usepackage{enumitem}
\usepackage{pifont}

\defcitealias{SFFT}{H22}

\shorttitle{SFFT for JWST}
\shortauthors{Hu \& Wang}

\begin{document}

\title{Differencing and Coadding JWST Images with Matched Point Spread Function}

\correspondingauthor{Lifan Wang}
\email{lifan@tamu.edu, leihu@andrew.cmu.edu}

\author[0000-0001-7201-1938]{Lei Hu}
\affiliation{Purple Mountain Observatory \\
Nanjing 210023, People's Republic of China}
\affiliation{McWilliams Center for Cosmology, Department of Physics, \\
Carnegie Mellon University, 5000 Forbes Ave, Pittsburgh, 15213, PA, USA}

\author[0000-0001-7092-9374]{Lifan Wang}
\affiliation{George P. and Cynthia Woods Mitchell Institute for Fundamental Physics \& Astronomy, \\
Texas A. \& M. University, Department of Physics and Astronomy, 4242 TAMU, College Station, TX 77843, USA}



\begin{abstract}

We present an algorithm to derive difference images for data taken with the JWST with matched point-spread functions (PSFs). It is based on the saccadic fast Fourier transform (SFFT) method but with revisions to accommodate the rotations and spatial variations of the PSFs. It allows for spatially varying kernels in B-spline form with separately controlled photometric scaling and \textit{Tikhonov} kernel regularization for harnessing the ultimate fitting flexibility. We present this method using the JWST/NIRCam images of galaxy cluster Abell 2744 acquired in JWST Cycle 1 as the test data. The algorithm can be useful for time-domain source detection and differential photometry with the JWST. It can also coadd images of multiple exposures taken at different field orientations. The coadded images preserve the sharpness of the central cores of the PSFs, and the positions and shapes of the objects are matched precisely with B-splines across the field. 

\end{abstract}



\section{Introduction} \label{sec:intro}

The James Webb Space Telescope (JWST) provides a unique opportunity for time-domain astronomy. The superb image quality enables the detection of faint transients out to the explosions of the first generation of massive stars and white dwarfs in the Universe \citep{Riess2006,FLARE2017,Regos2019,Lu2022}.
These diverse transients as different types of supernovae (SNe) can be direct probes to trace the cosmic star-formation history in the early Universe and expand our understanding substantially about the physics of the events at the epoch of the cosmic dawn.
Recent observations have demonstrated that very faint transients (mag $\sim$ 29) are abundant in the near-infrared images taken by the JWST \citep{Yan2023,Yan2023_2,2022TNSAN_GALSS_Chen,2022TNSAN_DDT_Chen,2022TNSAN_Hu,2023TNSAN_PEARLS_Yan,2023TNSAN_JADES17,2023TNSAN_GALSS_Chen}.

Image difference analysis is a major enabling technique in time-domain astronomy \citep{AL98,AL00,Bramich08,Miller08,HOTPANTS,SFFT}. However, for the JWST, the highly structured point spread function (PSF) and the uncertainties in precisely matching the astrometries of images taken at different epochs can pose significant challenges in the identification of transients in the vicinity of bright sources like the central regions of galaxies. These include, for example, strongly gravitationally lensed transients in the vicinity of foreground lensing galaxies \citep{DESILensing_Sheu2023} and the various nuclear transients including SNe as well as tidal disruption events \citep[e.g.,][]{SNeinAGN_Zhu2021,SNeinAGN_Grishin2021,TDE_Search_van2011,Regos2021}.
Calculating the difference images accurately is also important in high-precision differential photometry, which is employed in the detections of micro-lensing events \citep{Mao91,Sumi_2003,Sumi_2006,Sumi_2013} and exoplanet transits \citep[e.g.,][]{TESS_DIAPhot_Oelkers2018,TESS_DIAPhot_Montalto2020}. Recent works also reveal that the highly structured JWST/NIRCam PSF exhibits prominent spatial variation across the ﬁeld of view up to 20$\%$ and shows significant temporal variations at a level of $\sim$3$\%$-4$\%$ \citep{NIRCamPSF_Nardiello,Yan2023,NIRCamPSF_Zhuang1,NIRCamPSF_Zhuang2}.

This study presents a method based on the saccadic fast Fourier transform (SFFT) algorithm \citep[][hereafter \citetalias{SFFT}]{SFFT} to accommodate such complicated PSFs as those of the JWST for image differences. 
Unlike previously published algorithms, the SFFT can model the spatial variations of the PSF using a B-Spline function and allows for more accurate image matching across the field in terms of PSF homogenization and compensation of astrometric misalignments. 
By design, our algorithm presents image subtraction in Fourier space, thereby achieving exceptional computational efficiency. Moreover, it can be used for both sparse and crowded stellar fields. 
The SFFT has been applied and extensively examined in several ongoing time-domain surveys and some transient analyses \citep[e.g.,][]{TMTS2020,2022TNSAN_DESIRT,AST3-3Pipeline,Hu2017,Yang2022,Wang18evt,DESILensing_Sheu2023}. Recently, it also enabled new transient discoveries in JWST multi-epochs imaging observations \citep{2022TNSAN_Hu,2022TNSTR_AT2022aebo,2023TNSTR_AT2022aejk,2023TNSTR_AT2023ojj}.
The code of this study is built upon the SFFT algorithm proposed in \citetalias{SFFT} with significant improvements for the JWST. The improved version of SFFT is publicly available on Github\footnote{\url{https://github.com/thomasvrussell/sfft}}, and a tutorial demonstrating how to perform and evaluate SFFT subtraction on JWST/NIRCam is also provided\footnote{\url{https://github.com/thomasvrussell/sfft/blob/master/test/subtract_test_nircam/subtract4nircam.ipynb}}. 
The package is easily adaptable for data from other telescopes, such as with the Nancy Grace Roman Space Telescope \citep[Roman;][]{Roman} 
the Legacy Survey of Space and Time \citep[LSST;][]{LSST}.


\section{Test Data} \label{sec:testdata}

In this work, we demonstrate our method using the public JWST/NIRCam images of the Hubble Frontier Field (HFF) galaxy cluster Abell 2744 \citep{Abell2744_Castellano2016, Abell2744_Merlin2016}.
In Cycle 1, JWST has carried out imaging observations of the well-studied lensing cluster at multiple epochs by several JWST programs, which has built a great data set for testing image difference methods that can be used for JWST time-domain analyses.

We use the JWST/NIRCam imaging data acquired by the Early Release Science (ERS) GLASS-JWST program \citep[JWST-ERS-1324, PI Treu;][]{GLASS_Treu2022} on 2022 June 28-29 as the reference images of our subtraction tests. The data set consists of NIRCam images in seven filters (F090W, F115W, F150W, F200W, F277W, F356W, and F444W) covering observed wavelengths from 0.9 to 4.4 $\mu m$ with $5\sigma$ depths down to 28.8-29.7 AB magnitudes \citep[see][]{GLASS_Merlin2022}.

JWST revisited the cluster in November 2022 and conducted ultra-deep NIRCam observations with 4-6 hour exposures ($\sim$ 29-30 AB magnitudes) in seven filters (F115W, F150W, F200W, F277W, F356W, F410M, and F444W) as a part of another early JWST program that targets Abell 2744: the Ultradeep NIRSpec and NIRCam ObserVations before the Epoch of Reionization (UNCOVER) Treasury survey \citep[JWST-GO-2561, PI Labb\'e \& Bezanson;][]{UNCOVER_Bezanson2022}. 
We take the NIRCam observations collected by UNCOVER program on November 2, 2022, as the science images for subtractions. 
Note that the two JWST visits with different pointing and orientation overlap partially in a sky area covering $\sim$ 1.4 arcmin$^2$ centered at R.A. = $3^{\circ}_{\,\boldsymbol \cdot}5237919$, Decl. = $-30^{\circ}_{\,\boldsymbol \cdot}3673410$ \citep[see the JWST footprint of Abell 2744 in JWST Cycle 1 in][]{UNCOVER_Weaver2022}. All the JWST data used in this paper can be found in MAST: \dataset[10.17909/7qx3-zt80]{http://dx.doi.org/10.17909/7qx3-zt80}.

\section{Image Reduction and Mosaics} \label{sec:imgred}

We process the raw NIRCam data using the official STScI JWST Calibration Pipeline version 1.9.0\footnote{\url{https://github.com/spacetelescope/jwst}} \citep{jwst_pipeline_190} in the context of jwst\_1027.pmap\footnote{\url{https://jwst-crds.stsci.edu/context_table/jwst_1047.pmap}} that includes in-flight reference calibration files released on 2023 February 20. 
To enhance the reduction quality, we incorporate a few augmentations into the official pipeline largely following the data processing prescriptions of the Cosmic Evolution Early Release Science Survey \citep[CEERS; ][]{CEERS_Finkelstein2022} program described in \citet{CEERS_Bagley2023}

We reduce the uncalibrated images through Stage 1 (\texttt{Detector1Pipeline}) of the JWST Calibration Pipeline that performs detector-level corrections and converts ramps to count-rate (slope) images.
Given that the presence of stray light reﬂected off a secondary mirror support bar can introduce contamination to observations, giving rise to the characteristic ``wisp" features on images, we address this problem by undertaking the subtraction of wisp patterns on count-rate images in F150W and F200W (most prominent filters) using the available wisp templates released on 2022 August 26\footnote{\url{https://jwst-docs.stsci.edu/jwst-near-infrared-camera/nircam-features-and-caveats/nircam-claws-and-wisps}}.
Next, our processing turns to deal with the \textit{1/f noise} that is introduced during detectors read out \citep{Schlawin2020} and manifest itself in random horizontal and vertical striping patterns.
We identify and reduce the \textit{1/f noise} on count-rate images following the approach of amp-row and column subtraction proposed in \citet{CEERS_Bagley2023}.
Note that the stripes are measured after the sources in the field have been well masked, unlike \citet{CEERS_Bagley2023}, here we identify the source mask using \texttt{NoiseChisel} \citep{NoiseChisel}, a noise-based method tailored for the detection of very extended and diffuse objects. 
However, we note that the correction of \textit{snowballs} \citep{Rigby2023}, the circular defects on NIRCam images caused by cosmic-ray events, is yet to be included in our processing.
We perform additional instrumental corrections and calibrations (e.g., flat-fielding) by Stage 2 (\texttt{Image2Pipeline}) of the JWST Calibration Pipeline. In this step, the count-rate images are converted to units of $\textmd{MJy}\,\textmd{Sr}^{-1}$. 

We employ the JWST Stage 3 (\texttt{Image3Pipeline}) routine to create a mosaic image for each filter at the reference (science) epoch that combines all detectors and dithers with drizzling. 
This final stage consists of reduction steps including astrometric alignment (\texttt{TweakReg}), background matching (\texttt{SkyMatch}), outlier detection (\texttt{OutlierDetection}) and resampling (\texttt{Resample}).
As the \texttt{SkyMatch} step in Stage 3 may have difficulties in background matching for the cases with small dither \citep{CEERS_Bagley2023}, we skip this step in our processing but remove a single sky value for each individual image prior to Stage 3 using \texttt{SkyMatch} function\footnote{we adopt the mode with the parameter \texttt{skymethod} = ``local"}.
In addition, we opt to deactivate the \texttt{TweakReg} step in Stage 3, which is used to calculate the coordinate transformations for aligning individual images to an absolute World Coordinate System (WCS) frame. 
Instead, we first combine all detector images of each exposure to a single exposure image using the \texttt{Resample} function. We then select the exposure that has the maximal overlapping area with other exposures as the agent of mosaic creation to provide a reference WCS frame.
We use \texttt{SourceExtractor} \citep[][hereafter, SExtractor]{SExtractor} to create a source catalog for each exposure and perform relative astrometry with respect to the sky coordinates measured on the agent exposure using \texttt{SCAMP} \citep{SCAMP}. This step harmonizes the WCS information across all exposures involved in the mosaic creation without invoking any absolute WCS reference.
Upon running the \texttt{OutlierDetection} step to identify the outliers in the data, the exposures are drizzled to a single mosaic with a drizzling parameter \texttt{pixfrac} = 1 using JWST Stage 3 routine\footnote{Note that the mosaic images used for image subtractions are not significantly undersampled after the process of drizzling \citep[see][]{NIRCamPSF_Zhuang1}.}.

Finally, we make a custom sky subtraction on the mosaic by \texttt{NoiseChisel} to eliminate the residual background. 
Recall that the relative astrometry above is separately performed for each mosaic, thereby not guaranteeing WCS consistency across the mosaics. 
We then undertake additional relative astrometric calibrations at the mosaic level with respect to an agent mosaic in F200W.
The preprocessing ends with the final image resampling of all mosaics aligned to the agent mosaic using \texttt{SWarp} \citep{SWarp}. 
Throughout the paper, unless explicitly stated otherwise, the term ``mosaic" will refer to the astronomically aligned version after resampling.

\section{Image Subtraction} \label{sec:imgsub}

Our image subtraction method developed for optimal difference imaging of JWST data is based on the SFFT algorithm proposed in \citetalias{SFFT}.
Here, we briefly recap the algorithm framework and introduce the improvements in Section~\ref{subsec:sfft_method}. The specific subtraction scheme for JWST imaging observations is described in Section~\ref{subsec:jwst_subtract}.

\subsection{Improved SFFT Method} \label{subsec:sfft_method}

SFFT is an algorithm for astronomical image difference that presents the least-squares problem of image subtraction in Fourier space, bringing about a remarkable advancement in computational performance. 
It employs a $\delta$-function basis to allow for ultimately flexible image matching with ``shape-free" convolution kernels. 
Furthermore, SFFT can accommodate spatial variations of the matching kernel across the image field modeled by polynomials or B-splines.

For JWST observations, a number of modifications have been implemented within the SFFT framework. 
We summarize here:
\begin{itemize}
    \item The B-spline form spatial variation of convolution kernel, originally proposed in \citetalias{SFFT}, is now well integrated into our software and extensively tested on JWST imaging data.
    
    \item The photometric scaling factor through convolution, that is, the sum of the matching kernel, can be separately controlled, though it was fully entangled with the kernel pixels in \citetalias{SFFT}. For instance, the improved SFFT allows for a B-spline form matching kernel while imposing user-defined constraints on the kernel sum, such as being less flexible polynomials or constant across the field.
    
    \item The improved SFFT enables \textit{Tikhonov} regularization \citep{Press2007} to suppress the overfitting trend of matching kernel due to the ultimate flexibility of $\delta$-function basis. In particular, we provide an option to adjust our regularization to create a trivial penalty when the optimal matching kernel has a profile close to a $\delta$ function, e.g., for subtractions between images that already have similar PSFs.
\end{itemize}

\subsubsection{The SFFT: a recap} \label{subsubsec:sfft_framework}

The problem of image subtraction can be written as the minimization of the difference image $D$ \citep{AL98, AL00}, in the form
\begin{equation}
\begin{split}
    D(x, y) = S(x, y) & - (R \circledast K)(x, y) - B(x, y)
    \\
    = S(x, y) & - \iint dudv R(x-u, y-v)K_{x,y}(u, v) 
    \\
    & - B(x, y),
    \end{split}
    \label{eqn:sfft_eq1}
\end{equation}
where $R$ and $S$ are input images with same dimension $(N_0, N_1)$, $x$ and $y$ are image coordinate indices in the ranges of $[0, N_0)$ and $[0, N_1)$, respectively. The spatially varying convolution of image $R$ is denoted by $R \circledast K$.
$K_{x, y}$ is the matching kernel attached to image coordinate $(x, y)$ with shape $(2w_0+1, 2w_1+1)$, while $u$ and $v$ are kernel coordinate indices in the range of $[-w_0, w_0+1)$ and $[-w_1, w_1+1)$, respectively. 
Note that the matching kernel $K_{x, y}$, as a function of image coordinate, can vary across the image field to adapt to the ubiquitous spatial variations of PSF, photometric scaling and astrometry misalignment.
Furthermore, the additional offset map $B$ is used to account for the background difference between images $R$ and $S$.

Following \citet{Bramich08, Miller08}, SFFT decomposes the matching kernel $K_{x, y}$ into a complete $\delta$-function basis $\mathcal{K}$ as follows:
\begin{equation}
    K_{x,y} = \mathring{A}_{xy00} \mathcal{K}_{00} + \sum_{\alpha\beta} \mathring{A}_{xy\alpha\beta} \mathcal{K}_{\alpha\beta},
    \label{eqn:sfft_eq2}
\end{equation}
where $(\alpha, \beta)$ is the kernel coordinate of a non-center kernel pixel, i.e., $(\alpha, \beta) \neq \vec{0}$\footnote{By contrast, $(\alpha, \beta)$ included the center kernel pixel in \citetalias{SFFT}.}. 
The basis $\mathcal{K}$ is defined as,
\begin{equation}
    \label{eqn:sfft_eq3}
    \mathcal{K}_{00}(u, v) = \vec{\delta}(u, v)
\end{equation}
and
\begin{equation}
    \label{eqn:sfft_eq4}
    \begin{split}
    \mathcal{K}_{\alpha\beta}(u, v) = \vec{\delta}(u-\alpha, v-\beta) - \vec{\delta}(u, v),
    \end{split}
\end{equation}
where $\vec{\delta}$ is a binary function such that $\vec{\delta} (\rho, \epsilon) = 1$ if $\rho = \epsilon = 0$ and zero otherwise with $\rho$ and $\epsilon$ being any integers. 
Note that all basis vectors other than $\mathcal{K}_{00}$ have a zero-sum. As a result, the photometric scaling encapsulated in the convolution is uniquely determined by the coefficient $\mathring{A}_{xy00}$.

The kernel spatial variation is completely encoded in the coefficients $\mathring{A}_{xy00}$ (particular) and $\mathring{A}_{xy\alpha\beta}$ (general), which can be modeled by a two-dimensional smooth surface of either polynomials or B-splines across the image field:
\begin{equation}
    \label{eqn:sfft_eq5}
    \mathring{A}_{xy00} = \sum_{rs} \mathring{a}_{rs00} U_{rs}(x, y)
\end{equation}
and
\begin{equation}
    \label{eqn:sfft_eq6}
    \mathring{A}_{xy\alpha\beta} = \sum_{ij} \mathring{a}_{ij\alpha\beta} V_{ij}(x, y).
\end{equation}
The base functions $U_{rs}$ ($V_{ij}$) can have a $k$-order polynomial from:
\begin{equation}
    \label{eqn:sfft_eq7}
    U_{rs} (V_{ij}): \mathcal{P}_{\rho\epsilon}(x, y) = x^\rho y^\epsilon
\end{equation}
where $\rho$ and $\epsilon$ are polynomial power indices in the range of $[0, k]$ and $[0, k-\rho]$, respectively; Alternatively, they may follow a more flexible B-spline form:
\begin{equation}
    \label{eqn:sfft_eq8}
    U_{rs} (V_{ij}): \mathcal{B}_{\rho\epsilon}(x, y) = \mathbb{B}_{\rho; k, t_x}(x) \mathbb{B}_{\epsilon; k, t_y}(y),
\end{equation}
where $\mathbb{B}_{\rho; k, t_x}$ ($\mathbb{B}_{\epsilon; k, ty}$) are the one-dimensional B-spline basis functions of given degree $k$ and knots $t_x$ ($t_y$) along $x$ ($y$) axis, the indices $\rho$ and $\epsilon$ are in range of $[0, k+N_{t_x}]$ and $[0, k+N_{t_y}]$, respectively. 
We note that the improved SFFT, unlike in \citetalias{SFFT}, formulates the coefficients $\mathring{A}_{xy00}$ and $\mathring{A}_{xy\alpha\beta}$ independently. It signifies that the photometric scaling factor can be separately controlled, a feature that has been validated as beneficial in \citet{Bramich13}.

The differential background $B(x, y)$ is also fitted by a polynomial/B-spline form function:
\begin{equation}
    \label{eqn:sfft_eq9}
    B_{xy} = \sum_{pq} b_{pq} W_{pq}(x, y)
\end{equation}
where the base functions $W_{pq}$ are $\mathcal{P}_{pq}$ or $\mathcal{B}_{pq}$. 

With an approximation \citepalias[see Appendix~A of][]{SFFT} based on the fact that the scale of the spatial variations under consideration is significantly larger than that of convolution kernel, the Equation~(\ref{eqn:sfft_eq1}) can be rewritten as,
\begin{equation}
    \label{eqn:sfft_eq10}
    \begin{split}
    D(x, y) = & \,\, S(x, y) - \sum_{rs00} \mathring{a}_{rs00} U^{rs} \circ \mathcal{K}_{00}
    \\
    &- \sum_{ij\alpha\beta} \mathring{a}_{ij\alpha\beta} V^{ij} \circ \mathcal{K}_{\alpha\beta} - \sum_{pq} b_{pq} W^{pq},
    \end{split}
\end{equation}
where the notation $\circ$ indicates circular convolution and $U^{rs} = U_{rs} R$ and $V^{ij} = V_{ij} R$ for abbreviations.
The Equation~(\ref{eqn:sfft_eq10}) can be also derived using a different approach without the approximation (see the alternative perspective in Appendix~\ref{sec:appendix_approx}).

It is characteristic of the SFFT method to deliver the image subtraction problem forward into Fourier space. 
In the Fourier domain, we obtain
\begin{equation}
    \label{eqn:sfft_eq11}
    \begin{split}
    \widehat{D} = \widehat{S} &- \sum_{rs} a_{rs00} \widehat{U^{rs}}  \widehat{\mathcal{K}_{00}} - \sum_{ij\alpha\beta} a_{ij\alpha\beta} \widehat{V^{ij}} \widehat{\mathcal{K}_{\alpha\beta}}
    \\
    &- \sum_{pq} b_{pq}\widehat{W^{pq}}
    \end{split}
\end{equation}
where the symbols with a hat denote the Fourier transform of the images, $a_{rs00} = N_0N_1 \mathring{a}_{rs00}$ and $a_{ij\alpha\beta} = N_0N_1 \mathring{a}_{ij\alpha\beta}$.

By Parseval's theorem, the least-squares optimization of difference image $D$ is equivalent to minimizing its power spectrum in Fourier space. Let $G = \widehat{D} \widehat{D}^{*}$ ($^*$ stands for complex conjugate) that represents the power spectrum of the difference image, and we can define the \textit{loss} of the minimization in Fourier space as follows,
\begin{equation}
    \label{eqn:sfft_eq12}
    \mathcal{L}_0({\boldsymbol \theta}) = \chi^2 = \sum_{lm} G(l, m),
\end{equation}
where ${\boldsymbol \theta}$ are the free parameters ($a_{rs00}$, $a_{ij\alpha\beta}$ and $b_{pq}$) of Equation~\ref{eqn:sfft_eq11} that characterize the image subtraction. 
Thus, a least-square solution can be attained by optimizing its gradients to satisfy the condition $\nabla \mathcal{L}_0 = 0$.

\subsubsection{Kernel regularization} \label{subsubsec:sfft_regularization}

The minimization of image subtraction with $\delta$-function basis is prone to overfitting problems \citep{Becker12,Bramich16,iPTF_Masci17}. It is often characterized by irregularities (excessively noisy) resulting from matching kernels showing undesired adaptions to the noise of input data. The improved SFFT leverages \textit{Tikhonov} regularization to address this overfitting issue.

Following the prescription outlined in \citet{Becker12, Bramich16}, we regularize the shape of fitted matching kernels to have minimal local second derivatives by applying an additional penalty in the \textit{loss} function. 
Since SFFT has invariant matching kernels across the field, we must implement the regularization for the convolutional kernels realized at different positions.
With this consideration, we modify Equation~\ref{eqn:sfft_eq12} as follows,
\begin{equation}
    \label{eqn:sfft_eq13}
    \mathcal{L}({\boldsymbol \theta}) = \chi^2 + \lambda \sum_{g} w_g {\tilde{K}_g}^T L^T L \tilde{K}_g,
\end{equation}
where $\tilde{K}_g$ is the flatten version of matching kernel $K$ at image coordinate $(x_g, y_g)$. $\lambda$ is an empirically tuned parameter to adjust the overall strength of regularization. $w_g$ is the specific weight of regularization at the coordinate $(x_g, y_g)$. 
The specific weights accommodate the situations when some subregions may be more susceptible to overfitting so that a higher local suppression is required accordingly. For simplicity, we only use a uniform weighting scheme, i.e., $w_g = 1$, in this work.
$L$ is the symmetric $(N_K, N_K)$ \textit{Laplacian matrix} (also see \citet{Bramich16}) that represents the connectivity graph of kernel pixels, or equivalently, the standard Cartesian $\delta$-function basis, with elements
\begin{equation}
    \label{eqn:sfft_eq14}
    L (\mu, \nu) = \begin{cases}
      N_{\text{adj}, \mu}, & \text{for $\nu = \mu$, and $N_{\text{adj}}$ is the number} \\
        & \text{of kernel pixels adjacent to the} \\
        & \text{kernel pixel of index $\mu$.} \\
      -1, & \text{for $\nu \neq \mu$, and $\mu$ and $\nu$ correspond} \\
        & \text{to adjacent kernel pixels.} \\
      0, & \text{otherwise.} \\
    \end{cases}
\end{equation}
Note that $L \tilde{K}$ is an array of approximations to the local second derivative at each kernel pixel of $K$. It is locally calculated using at most five kernel pixels, generally following $f^{\prime\prime}(u, v) \approx f(u-1, v) + f(u+1, v) + f(u, v-1) + f(u, v+1) - 4 f(u, v)$ except for those kernel pixels adjacent to the boundary.

However, an important concern regarding the aforementioned definition lies in the fact that the regularization penalty remains non-trivial for a trivial matching kernel, i.e., a $\delta$-function kernel with its only non-zero element at the kernel center.
For example, the optimal matching kernels can be close to such a $\delta$-function when the image subtractions are performed on images with PSFs already broadly aligned to each other.
Under such circumstances, it is no longer reasonable to allow the regularization to hinder the image subtraction from finding a $\delta$-function-like kernel as its optimal solution.
Hence the improved SFFT offers users an option to remove the ``barrier" by further tweaking the central rows of the \textit{Laplacian matrix} as follows:
\begin{equation}
    \label{eqn:sfft_eq15}
    L (\mu_+, \nu) = 0
\end{equation}
where $\mu_+$ represents the central five kernel pixels of shape ``+" at $(-1, 0)$, $(1, 0)$, $(0, -1)$, $(0, 1)$ and $(0, 0)$. 
The modification is equivalent to dropping the contributions of the local second derivatives at these central kernel pixels from the summation of Equation~\ref{eqn:sfft_eq13}. As a result, the penalty is no longer a function of the pixel value of the kernel center at $(0, 0)$. In this work, we zero out the central rows of $L$ following Equation~\ref{eqn:sfft_eq15}.

Invoking Equations~\ref{eqn:sfft_eq2}-\ref{eqn:sfft_eq6} and using abbreviations $V_{ij}^g = V_{ij}(x_g, y_g)$ and $U_{rs}^g = U_{rs}(x_g, y_g)$, we flatten the matching kernel using the following equations
\begin{equation}
    \label{eqn:sfft_eq16}
    \tilde{K}_g(\eta) = \sum_{ij} \mathring{a}_{ij\alpha\beta} V_{ij}^g,
\end{equation}
and 
\begin{equation}
    \label{eqn:sfft_eq17}
    \tilde{K}_g(\phi) = \sum_{rs} \mathring{a}_{rs00} U_{rs}^g - \sum_{ij \alpha^\prime \beta^\prime} \mathring{a}_{ij \alpha^\prime \beta^\prime} V_{ij}^g,
\end{equation}
where $\eta$ is the flatten index of non-center kernel pixel $(\alpha, \beta)$, i.e., $\eta = (w_0+\alpha)(2w_1 + 1) + (w_1+\beta)$; $\phi$ is the flatten index of the center kernel pixel $(0, 0)$, i.e., $\phi = w_0(2w_1 + 1) + w1$;.
With taking regularization into account and abbreviation of $\mathsf{L} = L^T L$, the \textit{loss} function described in Equation~\ref{eqn:sfft_eq13} can be rewritten as
\begin{equation}
    \label{eqn:sfft_eq18}
    \begin{split}
    \mathcal{L}({\boldsymbol \theta}) = \chi^2 & + \lambda \sum_{g} w_g \{ \sum_{\eta^\prime \eta} \mathsf{L}_{\eta \eta^\prime} \tilde{K}_g(\eta) \tilde{K}_g({\eta^\prime})
    \\
    & + \sum_{\eta} \mathsf{L}_{\phi \eta} \tilde{K}_g({\phi}) \tilde{K}_g({\eta}) 
    \\
    & + \sum_{\eta} \mathsf{L}_{\eta \phi} \tilde{K}_g({\eta}) \tilde{K}_g({\phi}) + \mathsf{L}_{\phi \phi} {\tilde{K}_g({\phi})}^2 \}
    \end{split}
\end{equation}

\subsubsection{Subtraction in Fourier space} \label{subsubsec:sfft_subtract}

It remains to minimize of the \textit{loss} function $\mathcal{L}({\boldsymbol \theta})$ by optimizing its gradient with $\nabla \mathcal{L} = 0$, which still yields a linear system:
\begin{equation}
    \label{eqn:sfft_eq19}
    \mathbf{A} {\boldsymbol \theta} = \mathbf{b}.
\end{equation}
One can solve the linear equations following Appendix~\ref{sec:appendix_solve}.
Finally, taking the solution of Equation~\ref{eqn:sfft_eq19}, we arrive at the difference image $D$ by applying the solution to Equation~\ref{eqn:sfft_eq11}.

\subsection{Image Subtraction Scheme for JWST} \label{subsec:jwst_subtract}

This section presents the custom procedures of image subtractions we developed for JWST imaging data. 
To demonstrate our method, we perform the image subtraction between the NIRCam mosaic at the reference epoch (hereafter, reference mosaic) and the NIRCam mosaic at science epoch (hereafter, science mosaic) in each band (F115W, F150W, F200W, F277W, F356W, and F444W) created in Section~\ref{sec:imgred}.

\subsubsection{Cross convolution} \label{subsubsec:jwst_cross_conv}

Firstly, we convolve the reference (science) mosaic with the PSF model of science (reference) mosaic, i.e., the so-called cross-convolution. 
More specifically, the cross-convolved mosaics $R^*$ and $S^*$ are generated following
\begin{equation}
    \label{eqn:sfft_eq20}
    \begin{split}
    R^* &= R \circ P_S
    \\
    S^* &= S \circ P_R
    \end{split}
\end{equation}
where $R$ ($S$) is the reference (science) mosaic, and $P_R$ ($P_S$) is a corresponding PSF model at the image center retrieved from \texttt{WebbPSF} tool\footnote{\url{https://www.stsci.edu/jwst/science-planning/proposal-planning-toolbox/psf-simulation-tool}}. 

The image subtraction approach with cross-convolution involved to homogenize PSFs was initially proposed by \citet{Gal08} and \citet{Yuan08} and used in \citet{ZOGY}. 
Given that the JWST observations with a varying position angle lead to the rotation of the highly-structured PSF in the images taken at different epochs, a cross-convolution before a more sophisticated subtraction can broadly align the PSFs of input images to each other in a numerically stable way. 
As a result, it can effectively avoid deconvolution in the subsequent image subtraction that amplifies the noise and results in pathological correlation-induced patterns on the difference image \citep{ZOGY}.

Note that we do not construct the PSFs from the observed frames; the \texttt{WebbPSF} models are used instead as approximations to the PSFs of the observations. We do this because it is not always possible to construct PSFs from the observed images. The cross-convolution brings the images to a level of approximately matching PSFs. Further processing by SFFT will match the PSFs with the matching kernel given in Equation~\ref{eqn:sfft_eq2}.

\subsubsection{SFFT preprocessing} \label{subsubsec:jwst_sfft_preproc}

As described in \citetalias{SFFT}, it is essential for the SFFT method to solve the image subtraction on a masked version of input images rather than the original ones. A proper image mask serves to eliminate the affection of ``bad" pixels in real observations, e.g., saturation, cosmic rays, and variable sources. It thus directs the SFFT minimization towards an unbiased construction of convolution kernels.

In this work, we create a binary mask shared for the given science and reference mosaics, largely following the preprocessing routine of sparse-ﬂavor SFFT described in \citetalias{SFFT} with some modifications:
\begin{enumerate}
    \item We identify the ``bad" sources for SFFT subtraction using SExtractor. A field source is seen as ``bad'' if its SExtractor catalog value \texttt{FLAG} is zero and satisfies one of the following conditions: (1) it shows significant brightness change larger than one magnitude between reference and science mosaics measured by SExtractor catalog value \texttt{MAG\_AUTO}; (2) it is only detectable on one of the mosaic images.
    One may notice that the source exclusion here is a simplified version of the sparse-ﬂavor SFFT in \citetalias{SFFT}. We have skipped the identification of point-like sources as the deficiency of stars in JWST observations renders it difficult to detect a line feature regarding point sources using the Hough transformation.
    
    \item We create a binary mask initialized by ones, and then zeros out all pixels associated with the identified ``bad'' objects or background regions, according to SExtractor segmentation maps of reference and science mosaics.
    
    \item We tweak the binary mask by further clipping (converting mask value 1 to 0) all pixels that fall below three times of background standard deviation on any of the cross-convolved mosaics. The mask refinement is to make a better rejection of the noise-dominated pixels.
    Furthermore, our software allows for a final custom adjustment on the binary mask by clipping specific pixel regions defined in a SAOImage DS9 region file using polygon format.
    Although this function is not activated in our subtraction tests, it can be useful to exclude the saturated stars or other defects in the images when necessary.
\end{enumerate}
Note that the masked regions (mask value 1) are considered reliable for solving the image subtraction. To create the masked version of cross-convolved mosaics, we replace the pixels of cross-convolved mosaics located in unmasked regions (mask value 0) with trivial zeros.

\subsubsection{SFFT subtraction} \label{subsubsec:jwst_sfft_subtract}

We adopt the improved SFFT algorithm described in Section~\ref{subsec:sfft_method} to perform image subtraction between the cross-convolved mosaics produced in Section~\ref{subsubsec:jwst_cross_conv}.
This step aims to achieve better image matching to compensate the simple cross-convolution with limited matching accuracy.

We configure the tunable parameters of SFFT subtraction for JWST/NIRCam as follows:
\begin{itemize}
    \item The half-width of matching kernels is tuned to 11 and 5 pixels for short-wavelength and long-wavelength channels, respectively. 

    \item The general spatial variation of matching kernel (i.e., Equation~\ref{eqn:sfft_eq6}) is modeled by a flexible B-spline function following Equation~\ref{eqn:sfft_eq8} with degree two and knots uniformly distributed at regular intervals of 300 (150) pixels along both $x$ and $y$ axes for short-wavelength (long-wavelength) channels; 

    \item The particular spatial variation of matching kernel sum (i.e., Equation~\ref{eqn:sfft_eq5}) is fitted by a quadratic function following Equation~\ref{eqn:sfft_eq7}. 
    Note that a relatively low flexibility of kernel sum can preclude the photometric scaling from any undesired local structures. 
    Otherwise, for instance, a variable object not successfully masked in Section~\ref{subsubsec:jwst_sfft_preproc} could incur a local underestimate or overestimate of photometric scaling at the position of the variable source;

    \item Given that the background component of the mosaic images has been removed in Section~\ref{sec:imgred}, we assume a trivial flat differential background by a zero-order polynomial function (see Equation~\ref{eqn:sfft_eq9}). 
    
    \item The regularization strength parameter $\lambda$ (see Equation~\ref{eqn:sfft_eq13}) is finely tuned to be $3\times10^{-5}$. We use 512 positions randomly sampled within the image frame to regular the solution.
\end{itemize}

We solve the image matching on the masked version of cross-convolved mosaics created in Section~\ref{subsubsec:jwst_sfft_preproc}, and subsequently carry out the image subtraction between the (unmasked) cross-convolved mosaics by applying the matching solution\footnote{For a 2K NIRCam image, the kernel determination and image subtraction with B-splines typically take $\sim$10 seconds on one NVIDIA Tesla A100 GPU.}.
That is, 
\begin{equation}
    \label{eqn:sfft_eq21}
    \begin{split}
        D^* &= S^* - R^* \circledast K
    \end{split}
\end{equation}
where $K$ is the spatially variant matching kernel solved from the masked mosaics. An initial difference image $D^*$ is thus obtained. We hereafter refer to it as \textit{undecorrelated} difference.

\subsubsection{Noise Decorrelation} \label{subsubsec:jwst_noise_decorr}

Given that the \textit{undecorrelated} difference must possess highly correlated background noise introduced in the cross-convolution and SFFT subtraction, we whiten the difference image through a noise decorrelation procedure following \citetalias{SFFT} \citep[also see][]{ZOGY}. 

The formulation of noise decorrelation in \citetalias{SFFT} (see Appendix C of \citetalias{SFFT}) does not take kernel spatial variations into account, while the SFFT matching kernels are spatially variant across the image. 
We adopt a straightforward strategy by dividing the image frame into a grid of small tiles so that we can perform the noise decorrelation separately for each tile, where the matching kernel is approximately constant. 
Here we use a grid with tile size as small as the matching kernel size.
The noise decorrelation is performed by convolving undecorrelated difference with a spatially varying decorrelation kernel $Q$ that is constant for each tile in the grid: 
\begin{equation}
    \label{eqn:sfft_eq22}
    \begin{split}
        D &= D^* \circledast Q
    \end{split}
\end{equation}
and 
\begin{equation}
    \label{eqn:23}
    \begin{split}
    Q = \textbf{IDFT}(\sqrt{Z / ({\sigma_{S}^2 \abs{\widehat{P_R}}^2} + \sigma_{R}^2 \abs{\widehat{P_S}}^2 \abs{\widehat{K}}^2)})
    \end{split}
\end{equation}
where $\sigma_{R}$ ($\sigma_{S}$) is the background standard deviation of reference mosaic $R$ (science mosaic $S$), and $K$ is the mathcing kernel realized at each tile center. 
$Z$ is a factor that normalizes the decorrelation kernel $Q$ to have a unit kernel sum for preserving the flux zero-point. 
As the noise decorrelation is derived in Fourier space, the decorrelation kernel $Q$ is obtained after an inverse discrete Fourier transform (IDFT).
To avoid confusion, we refer to the final difference image $D$ as \textit{decorrelated} difference.

\subsubsection{Differential SNR} \label{subsubsec:jwst_diff_snr}

\begin{figure*}[ht!]
    \centering
    \includegraphics[trim=0cm 0cm 0cm 0cm,clip=true,width=17cm]{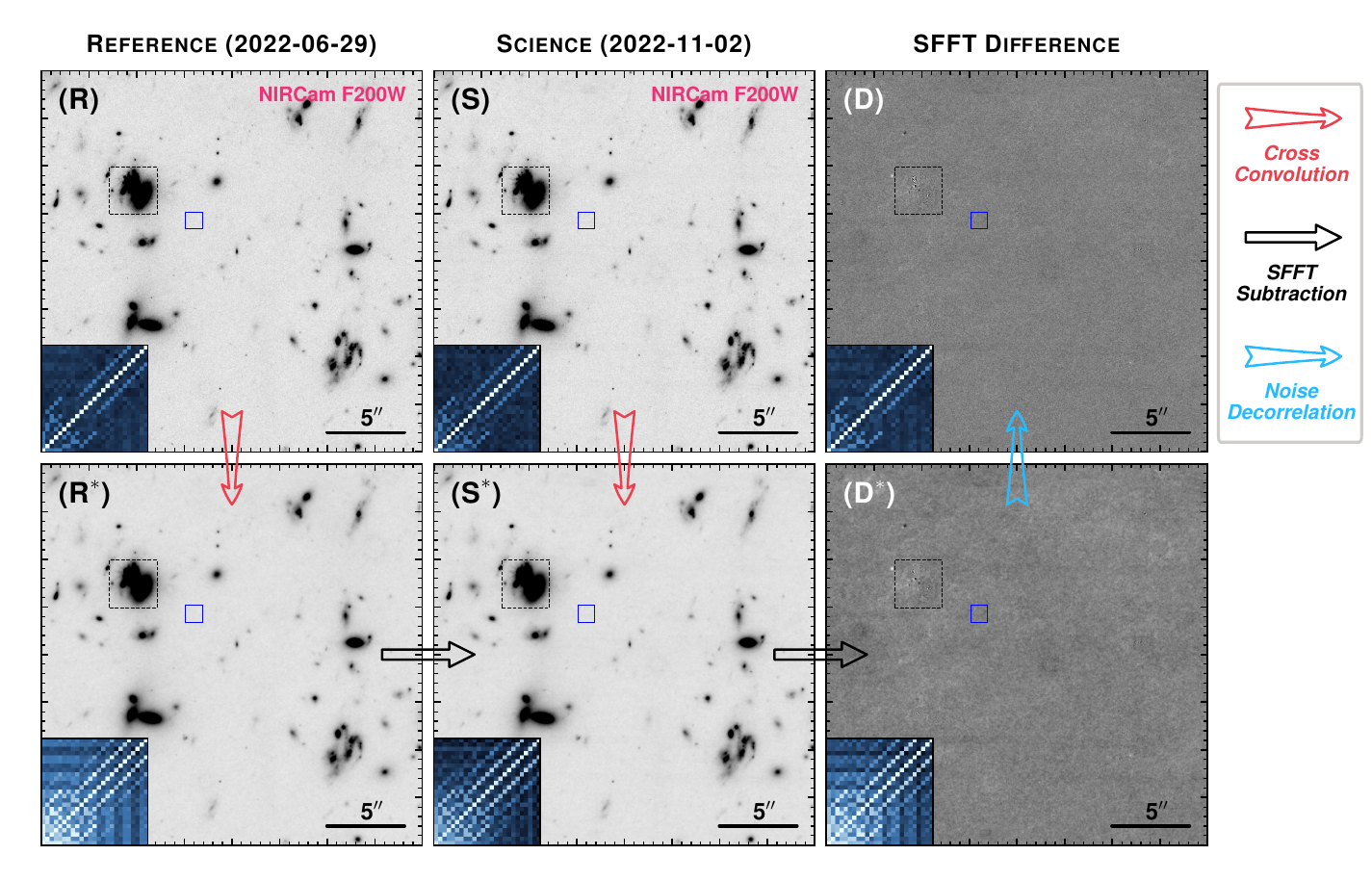}
    \caption{\label{fig:fig1} {Image subtraction performance of the SFFT method on NIRCam (F200W) mosaics of the Abell 2744 cluster.
    \textit{Top panels}: from left to right, reference mosaic (${\bf R}$), science mosaic (${\bf S}$), and images \textit{decorrelated} difference (${\bf D}$).
    \textit{Bottom panels}: from left to right, the cross-convolved reference mosaic (${\bf R^*}$), cross-convolved science mosaic (${\bf S^*}$), and \textit{undecorrelated} difference (${\bf D^*}$).
    The inset panel of each subplot shows the local noise correlation measured on the samples of the background pixels enclosed in the blue square. More specifically, it represents the covariance matrix of a multivariate random vector, consisting of 25 adjacent background pixels \citepalias[see definition in Appendix C of][]{SFFT}. The black dashed square indicates the position of transient candidate AT~2022acew. The arrows between different panels represent the corresponding operations as labeled on the right side.}}
\end{figure*}

\begin{figure*}[ht!]
    \centering
    \includegraphics[trim=0cm 0cm 0cm 0cm,clip=true,width=17cm]{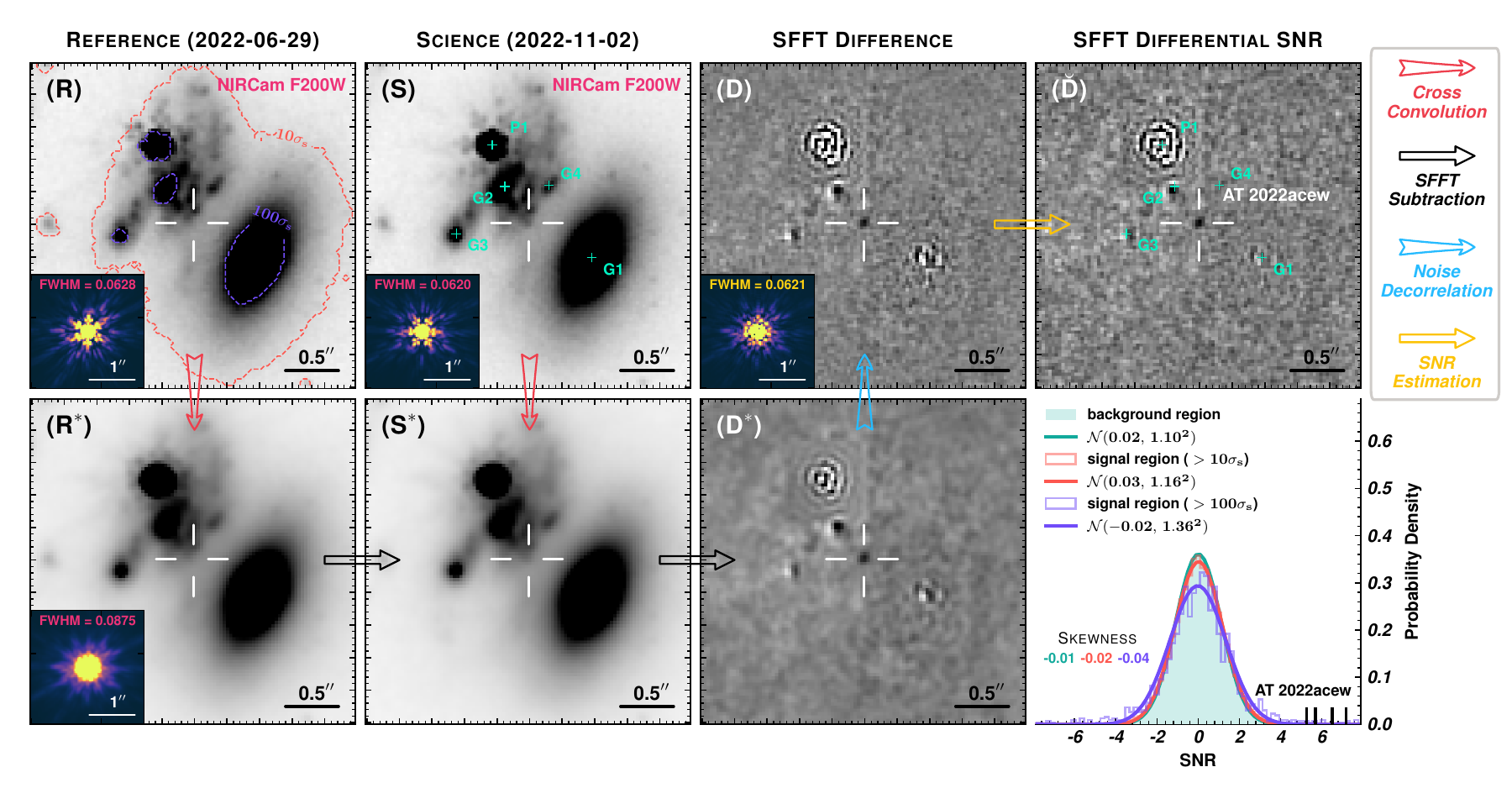}
    \caption{\label{fig:fig2} {Image subtraction performance of the SFFT method on NIRCam (F200W) mosaics in a close-up view around the transient candidate AT 2022acew (white cross) in the Abell 2744 cluster.  
    The grayscale images presented in the first three panel columns are the same as Figure~\ref{fig:fig1} but zoomed in a $3^{\prime\prime} \times 3^{\prime\prime}$ section around AT 2022acew. 
    However, the inset panels show the corresponding PSF models with measured FWHM values (in units of arcsec) labeled at the top. Note that the PSF models of ${\bf R}$ and ${\bf S}$ are retrieved from \texttt{WebbPSF}, while those of ${\bf R^*}$ and ${\bf D}$ are generated by applying the corresponding convolutions to original \texttt{WebbPSF} models.
    In the reference mosaic ${\bf R}$, the overlaid contour in red (purple) marks the detection profiles for pixels with flux over 10 (100) times the sky background standard deviation $\sigma_s$ on both reference mosaic ${\bf R}$ and science mosaic ${\bf S}$.
    The top rightmost panel shows the differential SNR map ${\bf \breve{D}}$, while the bottom rightmost panel presents the probability distributions (histograms in light colors) of differential SNRs within different regions across the image as labeled. 
    The cyan crosses mark the centroid positions of the five most prominent detections in this field measured by SExtractor on ${\bf S}$.
    The arrows between different image panels represent the corresponding operations as labeled on the right side. 
    A Gaussian profile is fitted for each distribution and illustrated by a dark-colored curve. The skewness of each distribution is marked in the corresponding color.
    The vertical black short lines at the right side of the x-axis indicate the differential SNR values of AT 2022acew higher than five.}}
\end{figure*}

A key objective of optimal image subtraction is to yield a difference image where the flux residues of non-variable sources and background are only dominated by their intrinsic statistical fluctuations, such as photon noise.
The least-squares minimization of SFFT subtraction serves this purpose by generally suppressing the flux residues as far as possible. However, the effort may also be overkill in this direction: significant overfitting can adapt to the noise in data and artificially drive the residues toward zeros. The concern motivates us to modulate the \textit{loss} function of SFFT subtraction with \textit{Tikhonov} regularization.
To evaluate whether the differential residues exactly stand at the optimal state, it is useful to derive the expected statistical noise of the difference image to provide a fiducial level of optimal subtraction.

Using a simple Monte Carlo sampling, we generate a propagated error map for \textit{decorrelated} difference.
JWST Stage 3 (\texttt{Image3Pipeline}) has produced an associated error map for each mosaic image. Firstly, we resample the JWST error map of the unaligned reference (science) mosaic to the target frame using \texttt{SWarp}.
Subsequently, we calibrate the resampled error map using a constant scaling factor so that the background errors can coincide with the actual flux distribution measured on the reference (science) mosaic.
Here, we have assumed that each pixel of reference (science) mosaic approximately follows an independent Gaussian distribution.

We randomly sample a zero-mean noise image 1024 times following the calibrated error map of reference (science) mosaic to trace the noise propagation through image subtraction. 
Next, we apply the same pixel operations involved in generating the \textit{decorrelated} difference (cross convolution, SFFT image subtraction, and decorrelation) to each randomly sampled noise image pair.
This step outputs 1024 propagated noise images at \textit{decorrelated} difference stage, and we calculate their standard deviation at each pixel to construct the final propagated error map. 

Finally, the signal-to-noise ratio (SNR) of \textit{decorrelated} difference is calculated as the \textit{decorrelated} difference $D$ divided by the propagated error map $N$:
\begin{equation}
    \label{eqn:sfft_eq24}
    \begin{split}
        \breve{D} &= D / N
    \end{split}
\end{equation}
hereafter referred to as differential SNR map.

The differential SNR map is a convenient check image to evaluate the quality of image subtraction. An optimal image subtraction should yield a differential SNR map that broadly adheres to an independent standard Gaussian distribution across the entire field.
However, this criterion is overly idealistic as the assumption of independent noise distribution of input reference (science) mosaic is only an approximation. 
Instead, we opt for a more pragmatic standard by considering the measured distribution of differential SNRs on the background as the fiducial level. 
One can judge the quality of image subtraction at sources by comparing its differential SNRs to the benchmark background level.

\section{Performance and Comparisons} \label{sec:performance_comparison}

In this section, we demonstrate our image differencing method described in Section~\ref{subsec:jwst_subtract} using NIRCam F200W mosaics of the Abell 2744 cluster created in Section~\ref{sec:imgred}.
We present the subtraction performance of our method in Section~\ref{subsec:subperf}. We compare our subtraction results with other methods in Section~\ref{subsec:comparison}.

\subsection{Subtraction Performance} \label{subsec:subperf}

Figure~\ref{fig:fig1} shows the subtraction performance of the SFFT method over a section of Abell 2744 covering $25^{\prime\prime} \times 25^{\prime\prime}$ centered at R.A. = $3^{\circ}_{\,\boldsymbol \cdot}5270483$, Decl. = $-30^{\circ}_{\,\boldsymbol \cdot}3645990$. 
Our SFFT method can subtract the JWST data with few conspicuous subtraction-induced artifacts in the \textit{decorrelated} difference, indicating an excellent image matching. 
The behavior of background noise correlation throughout the image subtraction process is depicted in Figure~\ref{fig:fig1}.
Initially, the reference (science) mosaic exhibits only weak local correlation. After the subsequent cross-convolution, there is a notable surge in noise correlation, which remains pronounced in the \textit{undecorrelated} difference.
However, our noise decorrelation efficiently reduces the prominent correlation to a level comparable to the original reference (science) mosaic.

Figure~\ref{fig:fig2} zooms in on a specific region of Abell 2744 centered at R.A. = $3^{\circ}_{\,\boldsymbol \cdot}5295458$, Decl. = $-30^{\circ}_{\,\boldsymbol \cdot}3648780$, where our subtraction method unveiled a transient candidate, AT~2022acew \citep{2022TNSAN_Hu}.
AT~2022acew appears at a sky position between two galaxies. One of the galaxies is a typical elliptical galaxy (SExtractor detection G1 in Figure~\ref{fig:fig2}), while the other might be a (lensed) galaxy that exhibits more intricate structures (SExtractor detections G2, G3, and G4 in Figure~\ref{fig:fig2}). 
Notably, a bright foreground star is near the galaxies (SExtractor detection P1 in Figure~\ref{fig:fig2}). 
This selected region showcases how our subtraction method performs on different sources with diverse morphology.
The differential SNR map in Figure~\ref{fig:fig2} reveals a distinct and prominent detection of AT~2022acew, despite its faint nature. 
Meanwhile, most pixels associated with the galaxies exhibit desirable low differential SNRs comparable to the background fiducial level. 
As shown in Figure~\ref{fig:fig2}, the subtraction quality on the central regions of bright objects G1, G3, and G4 have reached a (nearly) optimal level. For the object G2, our subtraction reveals structured residues at its core. Aperture photometry of G2 on the SFFT difference shows a significant positive net flux, which suggests that the residues are not likely a stand-alone subtraction-induced artifact, i.e., they may originate from the true flux variability of an AGN or a nuclear transient candidate at G2.
At the bright star P1, a circular subtraction artifact appears, but its contamination is well confined to a relatively small area with a radius of $0^{\prime\prime}_{\,\boldsymbol \cdot}22$ (equivalently, $\sim$3.5 times of FWHM) to the star's centroid.
We conduct a statistical analysis on the differential SNRs across the field presented in Figure~\ref{fig:fig2}.
It also confirms that the differential SNRs in signal-dominated regions are broadly consistent with the fiducial background level. 
Only a moderate performance deterioration appears at the brightest pixels (SNR $>$ 100). We emphasize that the subtraction quality on these regions is critical for the search and accurate photometry of the transients close to galaxy nuclei.
Figure~\ref{fig:fig2} also traces the PSF changes through the processes of image subtraction. One may notice that the noise decorrelation not only whitens the noise but also narrows down the PSF size of the difference image, rendering it comparable to original un-convolved mosaics.

\subsection{Comparisons with Other Subtraction Methods} \label{subsec:comparison}

\begin{figure*}[ht!] 
    \centering
    \includegraphics[trim=0cm 0cm 0cm 0cm,clip=true,width=17cm]{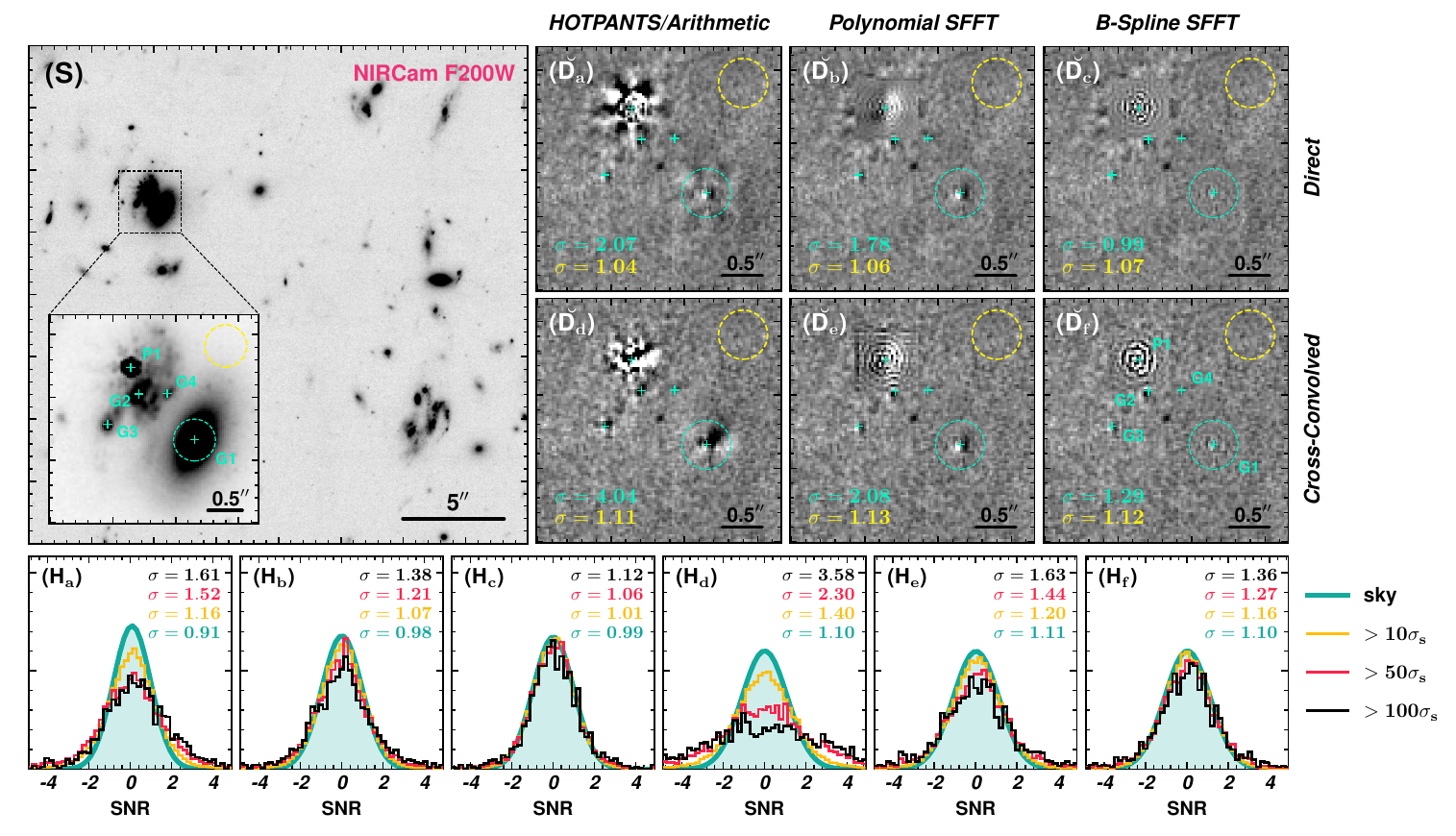}
    \caption{\label{fig:fig3} {Image subtraction performance on NIRCam (F200W) mosaics of the Abell 2744 cluster using different methods. ($\mathbf{\textbf{S}}$): A $25^{\prime\prime} \times 25^{\prime\prime}$ image section of science mosaic as shown in Figure~\ref{fig:fig1}, and its inset panel shows a $3^{\prime\prime} \times 3^{\prime\prime}$ zoomed-in postage stamp around AT 2022acew as shown Figure~\ref{fig:fig2}.
    ($\mathbf{\breve{\textbf{D}_a}}$) - ($\mathbf{\breve{\textbf{D}_f}}$): Postage stamps (with the same view as the inset panel of ${\bf S}$) of differential SNR maps from the six subtraction tests in Section~\ref{subsec:comparison}: direct HOTPANTS, direct polynomial SFFT, direct B-spline SFFT, cross-convolved arithmetic subtraction, cross-convolved polynomial SFFT, and cross-convolved B-spline SFFT, respectively.
    In each postage stamp, the cyan circle highlights the central region of the elliptical galaxy at the right corner, centered at R.A. = $3^{\circ}_{\,\boldsymbol \cdot}5293333$, Decl. = $-30^{\circ}_{\,\boldsymbol \cdot}3648317$, with a radius of $0^{\prime\prime}_{\,\boldsymbol \cdot}3098$, five times of FWHM size of the PSF model of science mosaic; The yellow circle marks a region selected from the ambient background with the same radius. 
    The standard deviation $\sigma$ of the pixel values enclosed by the cyan (yellow) circle is accordingly labeled at the left corner. The five prominent detections shown in Figure~\ref{fig:fig2} are also marked with cyan crosses in the inset panel of ${\bf S}$ and the panel $\mathbf{\breve{\textbf{D}_f}}$.
    ($\mathbf{{\textbf{H}_a}}$) - ($\mathbf{{\textbf{H}_f}}$): The histograms, from left to right, show the probability distributions of differential SNRs for different pixel regions for the six subtraction tests, respectively. 
    The pixel regions are identified following Figure~\ref{fig:fig2}. Each distribution's standard deviation $\sigma$ is labeled at the right border using the corresponding color.}}
\end{figure*}

\begin{figure*}[ht!] 
    \centering
    \includegraphics[trim=0.5cm 0cm 0cm 0cm,clip=true,width=18cm]{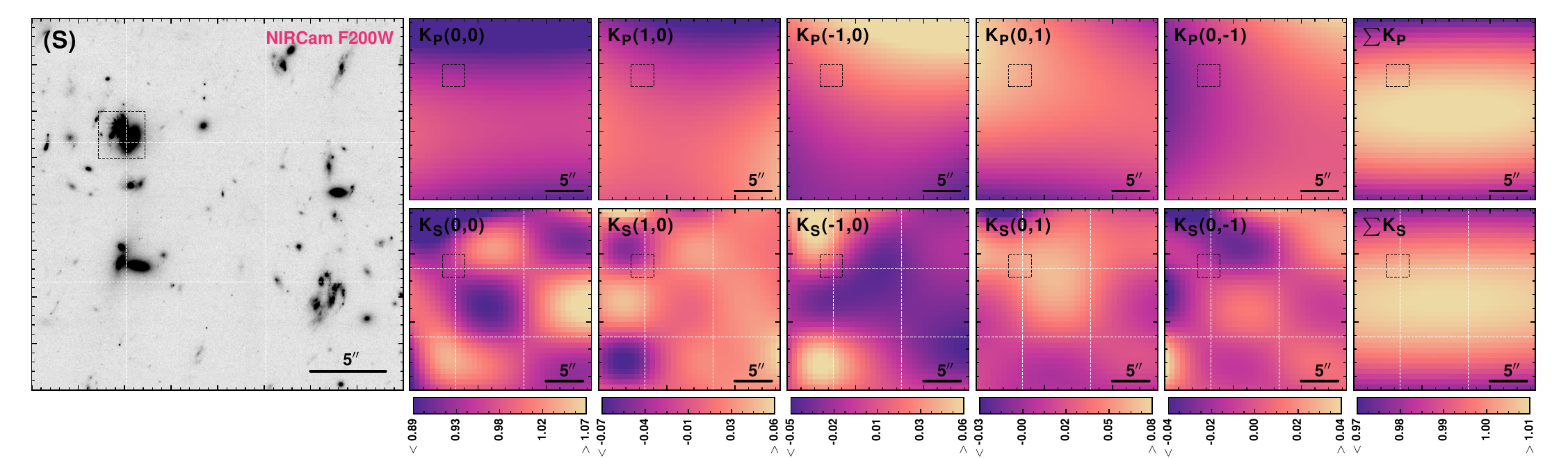}
    \caption{\label{fig:fig4} {Comparisons of spatial variations of matching kernels between the polynomial form SFFT and B-spline form SFFT. The leftmost panel shows the $25^{\prime\prime} \times 25^{\prime\prime}$ image section of science mosaic same as Figure~\ref{fig:fig1} and Figure~\ref{fig:fig2}}. The top-right panels that are labeled by $\textbf{K}_\textbf{P}(\alpha, \beta)$ show the spatial variations across the image field of the matching kernel pixel at $(\alpha, \beta)$ for cross-convolved polynomial SFFT test. In contrast, the top rightmost panel shows the spatial variations of the kernel sum. The bottom-right panels have the same style but for the cross-convolved B-spline SFFT test. In these panels, the dashed white lines indicate the positions of the inner knots of B-splines. The black dashed square indicates the position of AT 2022acew, same as Figure~\ref{fig:fig3}.}
\end{figure*}

To compare the performance of different subtraction methods, we conduct the following tests on NIRCam F200W mosaics of the Abell 2744 cluster using various approaches:
\begin{enumerate}[label=(\alph*)]
    \item Direct HOTPANTS: The subtraction is directly performed between science mosaic and reference mosaic using software {\tt\string HOTPANTS}\footnote{{\tt\string HOTPANTS} is configured with the following parameters: -c=t, -r=11, -ko=2, -bgo=0, -rss=30, -nsx=NX/100 and -nsy=NY/100.} \citep{HOTPANTS}, a widely used implementation of \citet{AL98} with Gaussian-function kernel basis. 
    Here, we use a simplified noise propagation, where the error map of the difference image is calculated as the square root of the variance sum of the input mosaics. Subsequently, we generate a corresponding differential SNR map based on this error map.

    \item Direct polynomial SFFT: The subtraction is directly carried out between science mosaic and reference mosaic using polynomial form SFFT. We create a binary mask following the steps described in Section~\ref{subsubsec:jwst_sfft_preproc} except the refinement clipping is applied to the un-convolved mosaics instead. 
    SFFT is configured with kernel spatial variation fitted by a quadratic function and inactivated \textit{Tikhonov} regularization, following Section~\ref{subsubsec:jwst_sfft_subtract} otherwise. 
    Like the direct HOTPANTS test, a differential SNR map is produced based on the simplified noise propagation.
    
    \item Direct B-spline SFFT: The subtraction is directly performed between science mosaic and reference mosaic using B-spline form SFFT. The same binary mask is used as in the direct polynomial SFFT test, and SFFT configurations completely follow Section~\ref{subsubsec:jwst_sfft_subtract}. Like the direct HOTPANTS test, a differential SNR map is generated using the simplified noise propagation.

    \item Cross-convolved arithmetic subtraction: The subtraction is performed between cross-convolved science mosaic and cross-convolved reference mosaic using a straightforward pixel-wise arithmetic operation of subtraction. 
    Subsequently, we whiten the correlated background noise of the resulting difference image introduced in cross-convolution following the noise decorrelation method outlined in Section~\ref{subsubsec:jwst_noise_decorr}. Finally, a differential SNR map is created using the Monte Carlo sampling described in Section~\ref{subsubsec:jwst_diff_snr}.

    \item Cross-convolved polynomial SFFT: The subtraction is conducted between cross-convolved science mosaic and cross-convolved reference mosaic using polynomial form SFFT, with the same configurations as the direct polynomial SFFT test.
    We proceed to decorrelate the background noise of the resulting difference image introduced in cross-convolution and SFFT subtraction following the strategy in Section~\ref{subsubsec:jwst_noise_decorr}. Again, we create a differential SNR map with the Monte Carlo sampling. 

    \item Cross-convolved B-spline SFFT: The subtraction test introduced in Section~\ref{subsec:jwst_subtract}.
\end{enumerate}

Figure~\ref{fig:fig3} shows the subtraction performances on NIRCam mosaics of the Abell 2744 cluster using the abovementioned methods. 
The statistics of differential SNR indicate that the two B-spline SFFT tests, with or without cross-convolution, have the best subtraction quality among the methods. Their differential SNRs in signal-dominated regions can broadly reach the fiducial background levels. 
It suggests that the introduction of flexible B-spline form spatial variations for convolutional kernels, coupled with the $\delta$-function kernel basis of SFFT, contributes to achieving a nearly optimal level of image subtraction for JWST/NIRCam data.
We note that the less optimal direct polynomial SFFT subtraction already demonstrates a considerable improvement over the direct HOTPANTS (see $\mathbf{{\textbf{H}_a}}$ and $\mathbf{{\textbf{H}_b}}$ in Figure~\ref{fig:fig3}). Despite both approaches adopting polynomial modeling for the PSF variation, SFFT's utilization of a $\delta$-function basis proves more effective for handling sophisticated PSF homogenization and compensation of astrometric misalignments for JWST, compared to {\tt\string HOTPANTS} with Gaussian basis functions. A detailed evaluation of polynomial form SFFT against other existing implementations (including {\tt\string HOTPANTS}) is presented in \citetalias{SFFT} with an emphasis on ground-based observation tests.

Considering the distributions of differential SNR alone, one may conclude that the direct B-spline SFFT appears to be as good as the cross-convolved B-spline SFFT.
However, it is important to emphasize that the inclusion of cross-convolution, in addition to its numerical stability, improves the subtraction quality for point sources.
The advantage is evident for the bright star P1: the direct B-spline SFFT gives rise to a more extended subtraction artifact, as shown by the 
residues surrounding the square-like pattern in Figure~\ref{fig:fig3}. 
Due to the limited number of stars in our test data, this advantage is not adequately reflected in the statistics.

To visualize the flexibility of the B-spline kernel, we have extracted the matching kernels at different image positions for the cross-convolved polynomial and B-spline SFFTs. Figure~\ref{fig:fig4} shows the spatial variations of five central kernel pixels and the kernel sum for these two subtraction tests.
Note that the photometric scaling for cross-convolved B-spline SFFT has been constrained to a low-degree quadratic function in the same form as cross-convolved polynomial SFFT. The two kernel sum surfaces depicted in Figure~\ref{fig:fig4} are roughly consistent with each other and close to a flat unit plane.
As anticipated, central kernel pixels for the B-spline form exhibit more structures across the field that can better adapt to the PSF spatial variation and compensate for local astrometrical misalignments.

\section{The Effect of Kernel Regularization} \label{sec:regeffect}

Overfitting is an important concern regarding any image subtraction method, especially with $\delta$-function basis. In this section, we demonstrate how the regularization technique assists the SFFT subtraction in properly harnessing the least-squares minimization process to mitigate overfitting.

\begin{figure*}[ht!] 
    \centering
    \includegraphics[trim=0cm 0cm 0cm 0cm,clip=true,width=17cm]{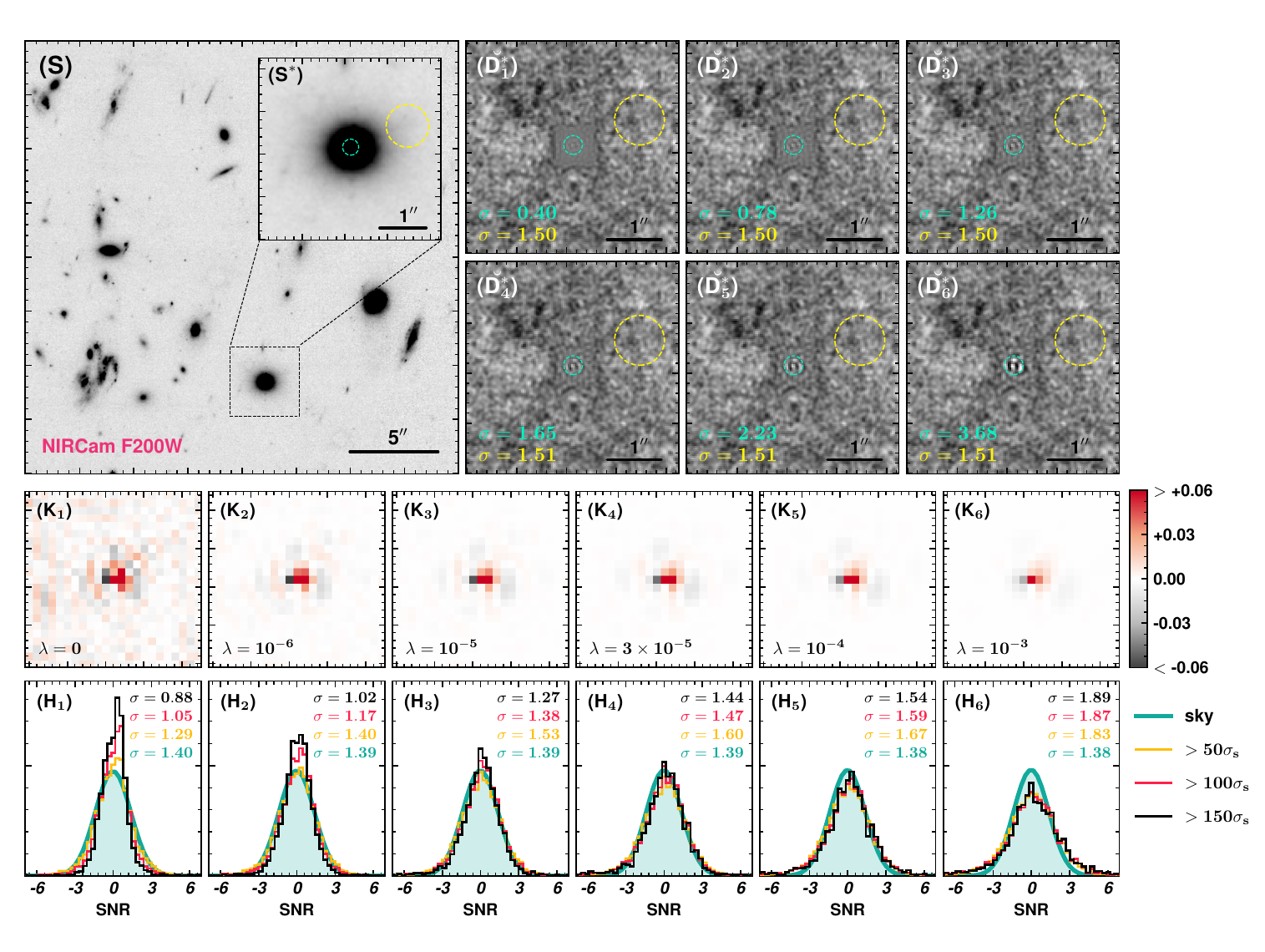}
    \caption{\label{fig:fig5} {The effect of \textit{Tikhonov} regularizations in SFFT subtraction. 
    ($\mathbf{\textbf{S}}$): A $25^{\prime\prime} \times 25^{\prime\prime}$ pixel section of science mosaic, centered at R.A. = $3^{\circ}_{\,\boldsymbol \cdot}5234850$, Decl. = $-30^{\circ}_{\,\boldsymbol \cdot}3615950$. 
    ($\mathbf{\textbf{S}^{*}}$): A $4^{\prime\prime} \times 4^{\prime\prime}$ postage stamp of the cross-convolved science mosaic centered at the examined galaxy.
    ($\mathbf{\breve{\textbf{D}^{*}_1}}$) - ($\mathbf{\breve{\textbf{D}^{*}_6}}$): Postage stamps (with a view same as ${\bf S^*}$) of the \textit{undecorrelated} differential SNR maps for SFFT subtraction tests with increasing regularization strength $\lambda$ being $0$, $10^{-6}$, $10^{-5}$, $3 \times 10^{-5}$, $10^{-4}$ and $10^{-3}$, respectively.
    In each postage stamp, the cyan circle highlights the central region of the examined galaxy with a radius of $0^{\prime\prime}_{\,\boldsymbol \cdot}1763$, twice the FWHM of the cross-convolved PSF model. In contrast, the yellow circle marks a region selected from the ambient background with a radius of 15 pixels ($0^{\prime\prime}_{\,\boldsymbol \cdot}4650$). 
    The standard deviation $\sigma$ of the pixel values enclosed by the cyan (yellow) circle is accordingly labeled at the left corner.
    ($\mathbf{\textbf{K}_1}$) - ($\mathbf{\textbf{K}_6}$): From left to right, the panels show the matching kernels realized at the position of the examined galaxy for SFFT subtraction tests with the increasing $\lambda$ as labeled.
    ($\mathbf{{\textbf{H}_1}}$) - ($\mathbf{{\textbf{H}_6}}$): From left to right, the histograms show the probability distributions of \textit{undecorrelated} differential SNRs, enclosed by different pixel regions shown as different histogram colors, for SFFT subtraction tests with the increasing $\lambda$. 
    The pixel regions are identified following Figure~\ref{fig:fig2} but defined on the cross-convolved image pair. Each distribution's standard deviation $\sigma$ is labeled at the right border in the corresponding color.}}
\end{figure*}

\begin{figure*}[ht!] 
    \centering
    \includegraphics[trim=0.7cm 0.5cm 0.5cm 0cm,clip=true,width=18cm]{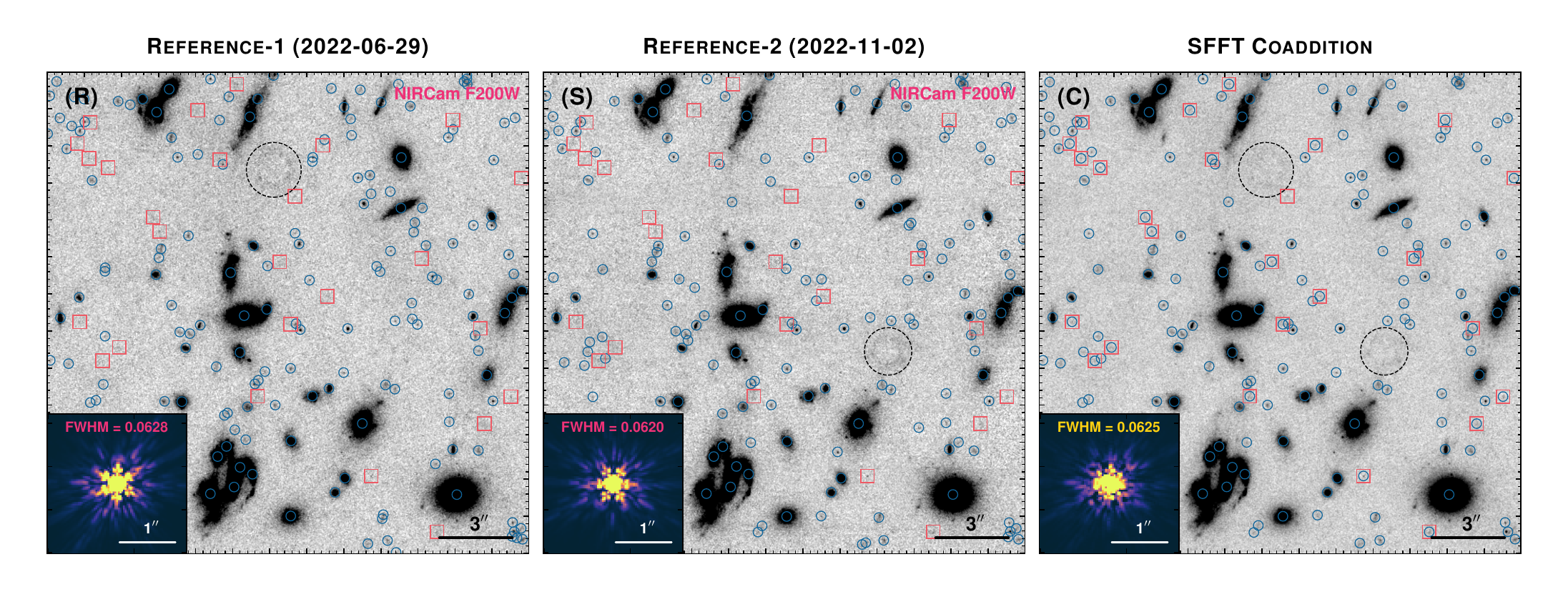}
    \caption{\label{fig:fig6} {Image co-add of NICam (F200W) mosaics of Abell 2744 cluster using the SFFT method. From left to right, the mosaic image was taken on 2022 Jun. 28-29, 2022 Nov. 2, and the SFFT co-add mosaic. The inset panels show the corresponding PSF models with measured FWHM values (in units of arcsec) labeled at the top. Note the PSF models of ${\bf R}$ and ${\bf S}$ are retrieved from \texttt{WebbPSF}, while the PSF models of co-added image ${\bf C}$ are generated by applying the corresponding convolutions to the \texttt{WebbPSF} models. In each panel, the blue circles are the sources detected by SExtractor on each panel image. The red squares represent the sources detected only in the co-added image. The SExtractor parameters were kept identical in detecting these sources. The black dashed circles indicate the pixel regions contaminated by the uncorrected \textit{snowballs} and sources inside them are not shown.}}
\end{figure*}

We reconduct our subtraction test on the NIRCam F200W mosaics six times, each with a different regularization strength $\lambda$ ranging from 0 to $10^{-3}$.
To evaluate the quality of \textit{undecorrelated} differences of these subtraction tests, we calculate a propagated error map for each \textit{undecorrelated} difference using the Monte Carlo sampling described in Section~\ref{subsubsec:jwst_diff_snr} and generate a corresponding \textit{undecorrelated} differential SNR map.

Figure~\ref{fig:fig5} shows the impact of \textit{Tikhonov} regularization on SFFT subtraction tests at an examined galaxy at R.A. = $3^{\circ}_{\,\boldsymbol \cdot}5216322$, Decl. = $-30^{\circ}_{\,\boldsymbol \cdot}3627870$ in Abell 2744. 
When no regularization is applied, one can notice that the overfitting manifests itself in the highly flattened \textit{undecorrelated} differential SNRs at the examined galaxy compared with the neighboring background.
Accordingly, the matching kernel at the examined galaxy is excessively noisy, signifying an undesirable adaptation to the galaxy's Poisson noise. 
Statistically, the overall distribution of \textit{undecorrelated} differential SNRs at signal-dominated regions is also pathological: it is even better than the fiducial background level.
As shown in Figure~\ref{fig:fig5}, adjusting the regularization strength can adequately address the overfitting problem. 
Increasing the regularization parameter $\lambda$ results in a smoother matching kernel and broadens the distribution of \textit{undecorrelated} differential SNRs in signal-dominated regions to be more reasonable. 
It is worth mentioning that the regularization strength we used in Section~\ref{subsec:jwst_subtract}, $\lambda = 3 \times 10^{-5}$, is a moderate value that effectively curbs overfitting without significantly compromising the quality of subtraction.

Despite its effectiveness in regularizing the noise levels of difference images, the \textit{Tikhonov} regularization may not be the ultimate solution to address the overfitting problem in image subtraction. The penalty term of \textit{Tikhonov} regularization only modulates the shape of the matching kernels but does not guarantee or properly quantify the optimal solution. It is not mathematically well-defined, and users must fine-tune the regularization strength as a hyper-parameter.
For optimal subtraction, the ultimate goal is to eliminate any structured residues in the difference image for all the objects without true variability. It is equivalent to minimizing the information content or maximizing the entropy of the residual images. 
In a future study, we will aim to incorporate a term to the loss function that can steer the fitting towards maximum entropy.

\section{Image Coadd} \label{subsec:coadd}

Naturally, the precise image matching accomplished by the B-spline form of SFFT can also be employed in the image co-addition to construct deeper mosaics. We illustrate this capability by performing a co-addition of the two NIRCam F200W mosaics of Abell 2744 that were used in Section~\ref{sec:performance_comparison}.

The co-addition scheme mostly inherits the procedures described in Section~\ref{subsec:jwst_subtract}. 
We use the same cross-convolution and B-spline form SFFT to align the input mosaics. The matched mosaics are subjected to weighted co-addition (instead of subtraction) followed by a noise decorrelation.
More specifically, we co-add the reference mosaic $R$ and science mosaic $S$ following
\begin{equation}
    \label{eqn:sfft_eq25}
    \begin{split}
        C = [w_S (S \circ P_R) + w_R (R \circ P_S) \circledast K] \circledast Q_c
    \end{split}
\end{equation}
and 
\begin{equation}
    \label{eqn:26}
    \begin{split}
    Q_c = \textbf{IDFT}(\sqrt{Z_c / ({{w_S^2 \sigma_{S}^2} \abs{\widehat{P_R}}^2} + {w_R^2 \sigma_{R}^2} \abs{\widehat{P_S}}^2 \abs{\widehat{K}}^2)})
    \end{split}
\end{equation}
where simple inverse variance weights $w_S = 1/\sigma_S^2$ and $w_R = 1/\sigma_R^2$ are adopted. Again, $Z_c$ is a normalization factor, and the noise decorrelation is performed over the same grid of tiles.
The example image co-add is shown in Figure~\ref{fig:fig6}. We note the sharp NIRCam PSF is preserved through the image co-addition. As expected, the co-add image bears reduced background noise and increased signal levels. With the detection parameters set to be identical \footnote{SExtractor is configured as following: \texttt{DETECT$\_$THERSH}=1.5, \texttt{DETECT$\_$MINAREA}=5 and \texttt{DEBLEND$\_$MINCONT}=0.005.}, the image co-add lead to detection limits that is $\sim$ 0.25 mag fainter, consistent with the expectation from Poisson statistics of the noise. The advantage of using SFFT for image co-adding is that it automatically aligns and matches the positions and profiles of the objects in the image field and creates co-add images that preserve the sharpness of the central cores of the JWST PSFs.

\section{Summary and Conclusions} \label{sec:conclude}

We have introduced an image differencing pipeline adapted to improve difference image analyses of JWST/NIRCam observations. We briefly summarize here the major steps of the pipeline: 
\begin{enumerate}
    \item Starting from uncalibrated NIRCam observations, we utilize an augmented version of the official STScI JWST Calibration Pipeline to create the reference and science mosaics taken at different epochs (see Section~\ref{sec:imgred}).
    
    \item Next, we perform cross-convolution by convolving the reference mosaic with the PSF model of the science mosaic, and vice versa, using the PSF models provided by \texttt{WebbPSF} (see Section~\ref{subsubsec:jwst_cross_conv}). This process can broadly align the PSFs of the two mosaics and minimize the numerical instability during the subsequent image subtraction. 
    
    \item We introduce a B-spline form of kernel variations in the SFFT method and modulate it by the \textit{Tikhonov} regularization to perform the image subtraction between the cross-convolved mosaics (see Section~\ref{subsubsec:jwst_sfft_preproc} and Section~\ref{subsubsec:jwst_sfft_subtract}).
    This step aims to achieve an accurate image matching with PSF homogenization and corrections of astrometrical misalignments. 
    This version of the SFFT method is characterized by several new features, outlined as follows:
    (1) It allows for flexible B-spline functions to depict the spatial variation of matching kernels, and in parallel, it can independently control the photometric scaling (kernel sum), typically by modeling with polynomial functions with lower degrees of freedom.
    (2) \textit{Tikhonov} regularization has been incorporated to suppress the undesired noise adaptions (i.e., overfitting problem) caused by the high level of flexibility in the fitting process. 
    The new method is detailed in Section~\ref{sec:imgsub}.
    
    \item We apply the prescription of noise decorrelation outlined in \citetalias{SFFT} to the difference image obtained from the SFFT subtraction (see Section~\ref{subsubsec:jwst_noise_decorr}). 
    This step effectively removes the convolution-induced correlations of background noise, and the FWHM of the central cores of the PSF becomes comparable to that in the original mosaics.
    
    \item Finally, the pipeline provides a differential SNR map as a check image to evaluate the quality of subtraction and diagnose possible overfitting (see Section~\ref{subsubsec:jwst_diff_snr}).
\end{enumerate}

This paper demonstrates the performance of the pipeline using JWST/NIRCam imaging data of the Abell 2744 cluster acquired in JWST Cycle 1 by the GLASS and UNCOVER programs. 
We exemplify that our method can achieve high subtraction quality, for which the residues on signal-dominated regions statistically harmonize with those at the fiducial background level. 
Moreover, we show the regularization technique can properly suppress the overfitting trend stemming from the high degree of freedom in SFFT subtraction.
We also make a comparison of subtraction performance using different techniques. Among them, a regularized B-spline form SFFT coupled with cross-convolution can achieve the best quality of image subtraction for the JWST/NIRCam data.
The method can also be used for accurate co-adding of JWST images. The algorithm is potentially useful in studying variable stars/transients in nearby galaxies, in searching for exoplanets through microlensing, and in finding SNe, especially those that are gravitationally lensed by nearby, relatively bright galaxies \citep[e.g.,][]{JWST_Cepheid_Yuan2022,JWST_Cepheid_Riess2023,JWST21aefx_Chen2023,WFIRST_Microlensing_Penny2019,LensedSN_Chen2022}.

\begin{figure*}[ht!] 
    \centering
    \includegraphics[trim=0.2cm 0cm 0cm 0cm,clip=true,width=18cm]{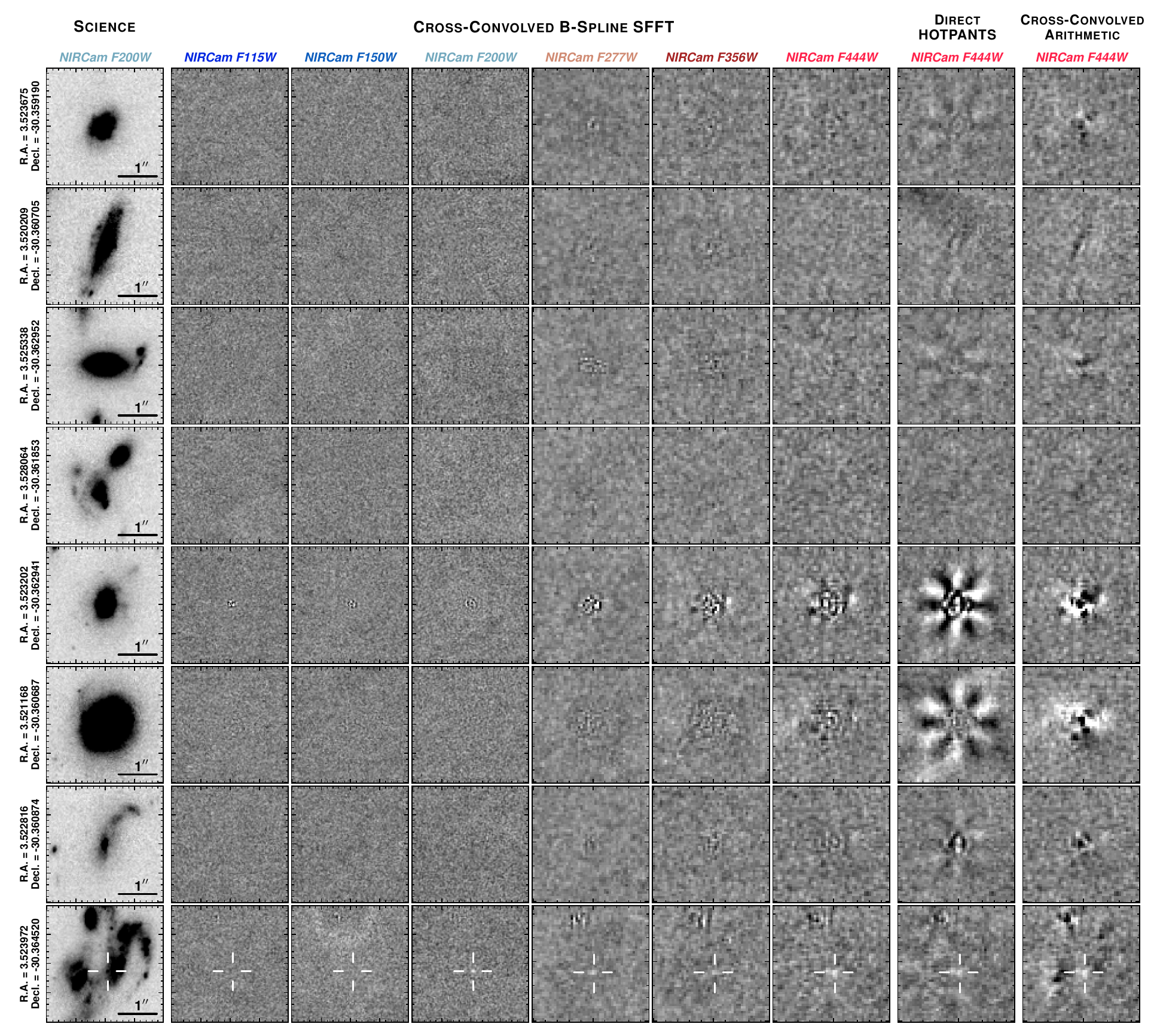}
    \caption{\label{fig:fig7} {Image subtraction performance of the SFFT method on NIRCam mosaics of Abell 2744 cluster illustrated with more examples. For each row, the first postage stamp shows a $3^{\prime\prime} \times 3^{\prime\prime}$ section of science mosaic in F200W centered at a sky coordinate as labeled on the left side. The next six postage stamps on the right-hand side present the differential SNR maps within the same view generated by the SFFT subtractions in F115W, F150W, F200W, F277W, F356W, and F444W, respectively. For comparisons, the last two postage stamps show the differential SNR maps generated by the direct HOTPANTS and the cross-convolved arithmetic subtractions (see Section~\ref{subsec:comparison}) in F444W, respectively.}}
\end{figure*}

Figure~\ref{fig:fig7} presents more examples of the subtraction performance on Abell 2744 cluster in different filter bands, zoomed in on several galaxies.
The data reduction and image subtraction follow the steps described in Section~\ref{sec:imgred} and Section~\ref{subsec:jwst_subtract}, respectively.
The PSF mismatch can lead to spurious features extending to $\gtrsim 0^{\prime\prime}.5$ from the center of a bright point source if the difference images are calculated using the original mosaics (See Figures~\ref{fig:fig3} and \ref{fig:fig7} for examples). While the total area affected by the PSF mismatch is usually insignificant compared to the entire image field, the central regions of galaxies require more careful treatment and can not be ignored for many important studies. For transient searches around diffuse galaxies, the effect of the PSF structure may not matter much. Our method is most important in searching for transients around galaxies with bright central cores. 
Some examples are the SNe and tidal disruption events close to the central regions of galaxies, AGN variabilities, and SNe lensed by foreground galaxies (whose Einstein ring is of the size $\sim 1^{\prime\prime}$). 

As shown in Figure~\ref{fig:fig7}, the difference images of these galaxies derived from our algorithm are clean and reveal no variabilities, whereas residual patterns are conspicuous when using simple arithmetic subtraction.  Our algorithm enables robust discoveries of variabilities and transients down to the nuclei of galaxies. The 8th row of Figure~\ref{fig:fig7} shows such an example (see the white cross).  
It shows a transient phenomenon that is significantly detected only at a wavelength longer than 2 $\mu m$, with magnitudes of $32.16\pm1.50$ (F115W), $31.27 \pm 0.73$ (F150W), $28.84 \pm 0.09$ (F200W),  $28.97 \pm 0.13$ (F277W), 
$ 29.13 \pm 0.16$ (F356W), and $28.27 \pm 0.10$ (F444W). This could be a highly reddened flare/TDE related to AGN variability, a very red SN, or a gravitationally magnified high redshift SN ($z \gtrsim 4$ for typical SNe~Ia around maximum light). Our pipeline facilitates the discoveries and follow-up studies of such objects to address these intriguing possibilities.

\begin{acknowledgments}

The authors thank the anonymous referee for the helpful comments that improved this paper. Lei Hu acknowledges support from the Jiangsu Funding Program for Excellent Postdoctoral Talent, China Postdoctoral Science Foundation (Grant No. 2022M723372), and the Major Science and Technology Project of Qinghai Province (2019-ZJA10). L. W. acknowledges support from the NSF grant AST-1817099. We thank Yanlong Hua and Xiangchong Li for useful discussion.

\end{acknowledgments}

%



\software{
    Astropy \citep{astropy13,astropy18,astropy22}, 
    SciPy \citep{2020SciPy-NMeth},
    Numpy \citep{2020NumPy-Array},
    MatPlotLib \citep{Matplotlib},
    CuPy \citep{CuPy},
    Scikit-image \citep{skimage},
    STScI JWST Calibration Pipeline \citep{jwst_pipeline_190},
    SourceExtractor \citep{SExtractor},
    SWarp \citep{SWarp},
    SCAMP \citep{SCAMP},
    HOTPANTS \citep{HOTPANTS},
    SFFT \citep{sfft_zenodo}
}



\clearpage
\appendix
\section{An alternative perspective on the approximation in SFFT}\label{sec:appendix_approx}

The Equation~\ref{eqn:sfft_eq10} can be obtained by tweaking the definition of the matching kernel $K$ as follows:
\begin{equation}
    \label{eqn:A1}
    \begin{split}
    K_{x,y}(0, 0) &= \sum_{rs} \mathring{a}_{rs00} U_{rs}(x, y) 
    \\
    &- \sum_{ij \alpha \beta} \mathring{a}_{ij \alpha \beta} V_{ij}(x, y),
    \end{split}
\end{equation}
and 
\begin{equation}
    \label{eqn:A2}
    K_{x,y}(\alpha, \beta) = \sum_{ij} \mathring{a}_{ij \alpha \beta} V_{ij}(x - \alpha, y - \beta).
\end{equation}
The center element $K_{x,y}(0, 0)$ aligns with the original definition in Equation~\ref{eqn:sfft_eq2}. The modification is applied to the non-center elements $K_{x,y}(\alpha, \beta)$, where we introduce a minor shift $(\alpha, \beta)$ on each basis function $V_{ij}$ of kernel spatial variation in the kernel construction. 
Consequently, this results in corresponding photometric scaling,
\begin{equation}
    \begin{split}       
    \label{eqn:A3}
    \sum K_{x, y} &= K_{x,y}(0, 0) + \sum_{\alpha\beta} K_{x,y}(\alpha, \beta) 
    \\
    &= \sum_{rs} \mathring{a}_{rs00} U_{rs}(x, y) 
    \\
    &- \sum_{ij \alpha \beta} \mathring{a}_{ij \alpha \beta} [V_{ij}(x, y) - V_{ij}(x - \alpha, y - \beta)].
    \end{split}
\end{equation}
We note that Equation~\ref{eqn:A1} and Equation~\ref{eqn:A2} give the exact form of the matching kernel performed in SFFT subtraction. 
In general, the basis functions $V_{ij}$ can be considered constant within the scale of matching kernel. With this consideration, the claimed matching kernel and photometric scaling in Section~\ref{subsubsec:sfft_framework} closely approximate their precise forms. For the sake of simplicity, we refrain from applying this refinement to the matching kernel throughout this paper.

\section{Solving Linear System in SFFT} \label{sec:appendix_solve}

The first component of the \textit{loss} function in Equation~\ref{eqn:sfft_eq18} represents the overall subtraction residues in the Fourier domain, calculated as the sum of the power spectrum of the difference image.
Substituting Equation~\ref{eqn:sfft_eq11} into Equation~\ref{eqn:sfft_eq12}, one can obtain the power spectrum
\begin{equation}
    \label{eqn:B1}
    \begin{split}
    G &= \widehat{S} \widehat{S}^* - 2 \sum_{rs} a_{rs00} \Re\;[\widehat{S}^* \widehat{U^{rs}}\widehat{\mathcal{K}_{00}}] 
    \\
    &- 2 \sum_{ij\alpha\beta} a_{ij\alpha\beta} \Re\;[\widehat{S}^* \widehat{V^{ij}}\widehat{\mathcal{K}_{\alpha\beta}}]
    \\
    &+2 \sum_{rs} \sum_{pq} a_{rs00}b_{pq} \Re\;[\widehat{U^{rs}} \widehat{\mathcal{K}_{00}} (\widehat{W^{pq}})^*]
    \\
    &+2 \sum_{ij\alpha\beta} \sum_{pq} a_{ij\alpha\beta}b_{pq} \Re\;[\widehat{V^{ij}} \widehat{\mathcal{K}_{\alpha\beta}} (\widehat{W^{pq}})^*]
    \\
    &+ \sum_{rs} \sum_{r^\prime s^\prime} a_{rs00} \widehat{U^{rs}} \widehat{\mathcal{K}_{00}} a_{r^\prime s^\prime 00} (\widehat{U^{r^\prime s^\prime}})^* (\widehat{\mathcal{K}_{00}})^*
    \\
    &+ \sum_{rs} \sum_{i^\prime j^\prime \alpha^\prime \beta^\prime} a_{rs00} \widehat{U^{rs}} \widehat{\mathcal{K}_{00}} a_{i^\prime j^\prime \alpha^\prime \beta^\prime} (\widehat{V^{i^\prime j^\prime}})^* (\widehat{\mathcal{K}_{\alpha^\prime \beta^\prime}})^* 
    \\
    &+ \sum_{ij\alpha\beta} \sum_{r^\prime s^\prime} a_{ij\alpha\beta} \widehat{V^{ij}} \widehat{\mathcal{K}_{\alpha\beta}} a_{r^\prime s^\prime 00} (\widehat{U^{r^\prime s^\prime}})^* (\widehat{\mathcal{K}_{00}})^* 
    \\
    &+ \sum_{ij\alpha\beta} \sum_{i^\prime j^\prime \alpha^\prime \beta^\prime} a_{ij\alpha\beta} \widehat{V^{ij}} \widehat{\mathcal{K}_{\alpha\beta}} a_{i^\prime j^\prime \alpha^\prime \beta^\prime} (\widehat{V^{i^\prime j^\prime}})^* (\widehat{\mathcal{K}_{\alpha^\prime \beta^\prime}})^* 
    \\
    &-2 \sum_{pq} b_{pq} \Re\;[\widehat{S} ^* \widehat{W^{pq}}] + \sum_{pq}\sum_{p^\prime q^\prime} b_{pq} \widehat{W^{pq}} b_{p^\prime q^\prime} (\widehat{W^{p^\prime q^\prime}})^*,
    \end{split}
\end{equation}
where $\Re$ stands for the real part of complex numbers.

Taking the regularization penalty into account, we can rewrite the \textit{loss} function described in Equation~\ref{eqn:sfft_eq18} by invoking the Equations~\ref{eqn:sfft_eq14},~\ref{eqn:sfft_eq16} and \ref{eqn:sfft_eq17}:
\begin{equation}
    \label{eqn:B2}
    \begin{split}
    \mathcal{L}({\boldsymbol \theta}) &= \sum_{lm} G(l, m) + \lambda \sum_{g} w_g \{ 
    \\  
    & \sum_{ij {i}^\prime{j}^\prime \eta {\eta}^\prime} [V_{ij}^g V_{{i}^\prime{j}^\prime}^g \mathsf{L}_{\eta {\eta}^\prime}] \mathring{a}_{ij \alpha\beta} \mathring{a}_{{i}^\prime{j}^\prime {\alpha}^\prime{\beta}^\prime}
    \\
    & + \sum_{{i}{j} rs {\eta}} [V_{{i}{j}}^g U_{rs}^g \mathsf{L}_{\phi {\eta}}] \mathring{a}_{{i}{j} {\alpha}{\beta}} \mathring{a}_{rs 00}
    \\
    & - \sum_{{i}^\prime{j}^\prime ij {\eta}^\prime{\eta}} [V_{{i}^\prime{j}^\prime}^g V_{ij}^g \mathsf{L}_{\phi {\eta}}] \mathring{a}_{{i}^\prime{j}^\prime {\alpha}^\prime{\beta}^\prime} \mathring{a}_{ij \alpha\beta}
    \\
    & + \sum_{ij {r}{s} \eta} [V_{ij}^g U_{{r}{s}}^g \mathsf{L}_{\eta \phi}] \mathring{a}_{ij \alpha\beta} \mathring{a}_{{r}{s} 00}
    \\
    & - \sum_{{i}^\prime{j}^\prime ij {\eta}^\prime{\eta}} [V_{{i}^\prime{j}^\prime}^g V_{ij}^g \mathsf{L}_{{\eta} \phi}] \mathring{a}_{{i}^\prime{j}^\prime {\alpha}^\prime{\beta}^\prime} \mathring{a}_{ij \alpha\beta}
    \\
    & + \sum_{rs {r}^\prime{s}^\prime} [U_{rs}^g U_{{r}^\prime{s}^\prime}^g \mathsf{L}_{\phi \phi}] \mathring{a}_{rs 00} \mathring{a}_{{r}^\prime{s}^\prime 00}
    \\
    & + \sum_{ij {i}^\prime{j}^\prime \eta {\eta}^\prime} [V_{ij}^g V_{{i}^\prime{j}^\prime}^g \mathsf{L}_{\phi \phi}] \mathring{a}_{ij \alpha\beta} \mathring{a}_{{i}^\prime{j}^\prime {\alpha}^\prime{\beta}^\prime}
    \\
    & -2 \sum_{rs {i}{j} {\eta}} [U_{rs}^g V_{{i}{j}}^g \mathsf{L}_{\phi \phi}] \mathring{a}_{rs 00} \mathring{a}_{{i}{j} {\alpha}{\beta}} \},
    \end{split}
\end{equation}
where $\eta$ and ${\eta}^\prime$ are the flatten index of non-center kernel pixels $(\alpha, \beta)$ and $(\alpha^\prime, \beta^\prime)$, respectively. $\phi$ is the flatten index of the center kernel pixel $(0, 0)$.

Next, we optimize the gradient of the \textit{loss} function with $\nabla \mathcal{L} = 0$ that constructs the linear system described in Equation~\ref{eqn:sfft_eq19}. 
By virtue of the simple form of $\delta$-function basis in Fourier space, we can write the linear system as follows. 

The matrix $\mathbf{A}$ is symmetric and can be seen as a partitioned matrix with four submatrices.
The upper left block of $\mathbf{A}$: (\textbf{i}) the component of the cross terms for ``kernel to kernel",
\begin{equation}
    \label{eqn:B3}
    \begin{split}
    \mathbf{A}(\bar{i}\bar{j} \bar{\alpha}\bar{\beta}, ij\alpha\beta) = &-\Omega_{\bar{i}\bar{j} ij}(\bar{\alpha}, \bar{\beta}) - \Omega_{\bar{i}\bar{j} ij} (-\alpha, -\beta) 
    \\
    & + \Omega_{\bar{i}\bar{j} ij} (\bar{\alpha} - \alpha, \bar{\beta} - \beta) + \Omega_{\bar{i}\bar{j} ij} (0, 0)
    \\
    & + \lambda \sum_{g} w_g V_{\bar{i}\bar{j}}^g V_{ij}^g (\mathsf{L}_{\bar{\eta} \eta} + \mathsf{L}_{\eta \bar{\eta}} - \mathsf{L}_{\phi \bar{\eta}}
    \\
    & - \mathsf{L}_{\phi \eta} - \mathsf{L}_{\bar{\eta} \phi} - \mathsf{L}_{\eta \phi} + 2 \mathsf{L}_{\phi \phi})/N_0^2N_1^2,
    \end{split}
\end{equation}
where
\begin{equation}
    \label{eqn:B4}
    \begin{split}
    \Omega_{\bar{i}\bar{j} ij} (\rho, \epsilon) &= \mathring{\Omega}_{\bar{i}\bar{j} ij} (\rho\bmod{N0}, \epsilon\bmod{N1})
    \\
    \mathring{\Omega}_{\bar{i}\bar{j} ij} &= \frac{1}{N_0N_1} \Re\;[\textbf{DFT}(\widehat{V^{\bar{i}\bar{j}}} (\widehat{V^{ij}})^*)], 
    \end{split}
\end{equation}
where, following the previous convention, $\bar{\eta}$ is the flatten index of non-center kernel pixels $(\bar{\alpha}, \bar{\beta})$;
(\textbf{ii}) the component of the cross terms for ``kernel to scaling",
\begin{equation}
    \label{A5}
    \begin{split}
    \mathbf{A}(\bar{i}\bar{j} \bar{\alpha}\bar{\beta}, rs 00) = &\,\,\Omega_{\bar{i}\bar{j} rs} (\bar{\alpha}, \bar{\beta}) - \Omega_{\bar{i}\bar{j} rs} (0, 0)
    \\
    & + \lambda \sum_{g} w_g V_{\bar{i}\bar{j}}^g U_{rs}^g (\mathsf{L}_{\phi \bar{\eta}} + \mathsf{L}_{\bar{\eta} \phi}
    \\
    & - 2 \mathsf{L}_{\phi \phi})/N_0^2N_1^2,
    \end{split}
\end{equation}
where
\begin{equation}
    \label{A6}
    \begin{split}
    \Omega_{\bar{i}\bar{j} rs} (\rho, \epsilon) &= \mathring{\Omega}_{\bar{i}\bar{j} rs} (\rho\bmod{N0}, \epsilon\bmod{N1})
    \\
    \mathring{\Omega}_{\bar{i}\bar{j} rs} &= \frac{1}{N_0N_1} \Re\;[\textbf{DFT}(\widehat{V^{\bar{i}\bar{j}}} (\widehat{U^{rs}})^*)];
    \end{split}
\end{equation}
(\textbf{iii}) the component of the cross terms for ``scaling to kernel",
\begin{equation}
    \label{A7}
    \begin{split}
    \mathbf{A}(\bar{r}\bar{s} 00, ij\alpha\beta) = &\,\,\Omega_{\bar{r}\bar{s} ij} (-\alpha, -\beta) - \Omega_{\bar{r}\bar{s} ij} (0, 0)
    \\
    & + \lambda \sum_{g} w_g U_{\bar{r}\bar{s}}^g V_{ij}^g 
    (\mathsf{L}_{\phi \eta} + \mathsf{L}_{\eta \phi}
    \\
    &- 2 \mathsf{L}_{\phi \phi})/N_0^2N_1^2,
    \end{split}
\end{equation}
where
\begin{equation}
    \label{A8}
    \begin{split}
    \Omega_{\bar{r}\bar{s} ij} (\rho, \epsilon) &= \mathring{\Omega}_{\bar{r}\bar{s} ij} (\rho\bmod{N0}, \epsilon\bmod{N1})
    \\
    \mathring{\Omega}_{\bar{r}\bar{s} ij} &= \frac{1}{N_0N_1} \Re\;[\textbf{DFT}(\widehat{U^{\bar{r}\bar{s}}} (\widehat{V^{ij}})^*)];
    \end{split}
\end{equation}
(\textbf{iv}) the component of the cross terms for ``scaling to scaling",
\begin{equation}
    \label{A9}
    \begin{split}
    \mathbf{A}(\bar{r}\bar{s} 00, rs 00) = &\,\,\Omega_{\bar{r}\bar{s} rs} (0, 0)
    \\
    & + \lambda \sum_{g} w_g U_{\bar{r}\bar{s}}^g U_{rs}^g 2 \mathsf{L}_{\phi \phi}/N_0^2N_1^2,
    \end{split}
\end{equation}
where
\begin{equation}
    \label{A10}
    \begin{split}
    \Omega_{\bar{r}\bar{s} rs} (\rho, \epsilon) &= \mathring{\Omega}_{\bar{r}\bar{s} rs} (\rho\bmod{N0}, \epsilon\bmod{N1})
    \\
    \mathring{\Omega}_{\bar{r}\bar{s} rs} &= \frac{1}{N_0N_1} \Re\;[\textbf{DFT}(\widehat{U^{\bar{r}\bar{s}}} (\widehat{U^{rs}})^*)].
    \end{split}
\end{equation}

The upper right block of $\mathbf{A}$: (\textbf{i}) the component of the cross terms for ``kernel to background",
\begin{equation}
    \label{A11}
    \mathbf{A}(\bar{i}\bar{j} \bar{\alpha}\bar{\beta}, pq) = \Lambda_{\bar{i}\bar{j} pq}(\bar{\alpha}, \bar{\beta}) - \Lambda_{\bar{i}\bar{j} pq} (0, 0),
\end{equation}
where
\begin{equation}
    \label{A12}
    \begin{split}
    \Lambda_{\bar{i}\bar{j} pq} (\rho, \epsilon) &= \mathring{\Lambda}_{\bar{i}\bar{j} pq} (\rho\bmod{N_0}, \epsilon\bmod{N_1})
    \\
    \mathring{\Lambda}_{\bar{i}\bar{j} pq} &= \Re\;[\textbf{DFT}(\widehat{V^{\bar{i}\bar{j}}} (\widehat{W^{pq}})^*];
    \end{split}
\end{equation}
(\textbf{ii}) the component of the cross terms for ``scaling to background",
\begin{equation}
    \label{A13}
    \mathbf{A}(\bar{r}\bar{s} 00, pq) = \Lambda_{\bar{r}\bar{s} pq} (0, 0),
\end{equation}
where
\begin{equation}
    \label{A14}
    \begin{split}
    \Lambda_{\bar{r}\bar{s} pq} (\rho, \epsilon) &= \mathring{\Lambda}_{\bar{r}\bar{s} pq} (\rho\bmod{N_0}, \epsilon\bmod{N_1})
    \\
    \mathring{\Lambda}_{\bar{r}\bar{s} pq} &= \Re\;[\textbf{DFT}(\widehat{U^{\bar{r}\bar{s}}} (\widehat{W^{pq}})^*].
    \end{split}
\end{equation}

The lower left block of $\mathbf{A}$: (\textbf{i}) the component of the cross terms for ``background to kernel",
\begin{equation}
    \label{A15}
    \mathbf{A}(\bar{p}\bar{q}, ij \alpha\beta) = \Psi_{\bar{p}\bar{q} ij}(-\alpha, -\beta) - \Psi_{\bar{p}\bar{q} ij} (0, 0),
\end{equation}
where
\begin{equation}
\label{A16}
\begin{split}
\Psi_{\bar{p}\bar{q} ij} (\rho, \epsilon) &= \mathring{\Psi}_{\bar{p}\bar{q} ij} (\rho\bmod{N_0}, \epsilon\bmod{N_1})
\\
\mathring{\Psi}_{\bar{p}\bar{q} ij} &= \Re\;[\textbf{DFT}(\widehat{W^{\bar{p}\bar{q}}} (\widehat{V^{ij}})^*)];
\end{split}
\end{equation}
(\textbf{ii}) the component of the cross terms for ``background to scaling",
\begin{equation}
    \label{A17}
    \mathbf{A}(\bar{p}\bar{q}, rs 00) = \Psi_{\bar{p}\bar{q} rs} (0, 0),
\end{equation}
where
\begin{equation}
\label{A18}
\begin{split}
\Psi_{\bar{p}\bar{q} rs} (\rho, \epsilon) &= \mathring{\Psi}_{\bar{p}\bar{q} rs} (\rho\bmod{N_0}, \epsilon\bmod{N_1})
\\
\mathring{\Psi}_{\bar{p}\bar{q} rs} &= \Re\;[\textbf{DFT}(\widehat{W^{\bar{p}\bar{q}}} (\widehat{U^{rs}})^*)].
\end{split}
\end{equation}

The lower right block of $\mathbf{A}$ comprised of the cross terms for ``background to background":
\begin{equation}
    \label{A19}
    \begin{split}
    \mathbf{A}_{\bar{p}\bar{q} pq} &= \Phi_{\bar{p}\bar{q} pq} (0, 0),
    \end{split}
\end{equation}
where
\begin{equation}
    \label{A20}
    \begin{split}
    \Phi_{\bar{p}\bar{q} pq} (\rho, \epsilon) &= \mathring{\Phi}_{\bar{p}\bar{q} pq} (\rho\bmod{N_0}, \epsilon\bmod{N_1})
    \\
    \mathring{\Phi}_{\bar{p}\bar{q} pq} &= N_0N_1 \Re\;[\textbf{DFT}(\widehat{W^{\bar{p}\bar{q}}} (\widehat{W^{pq}})^*)]
    \end{split}
\end{equation}

One the right hand of the linear system, the upper block of $\mathbf{b}$: (\textbf{i}) the component for ``kernel",
\begin{equation}
    \label{A21}
    \mathbf{b}_{\bar{i}\bar{j} \bar{\alpha}\bar{\beta}} = \Theta_{\bar{i}\bar{j}}(\bar{\alpha}, \bar{\beta}) - \Theta_{\bar{i}\bar{j}} (0, 0)
    \end{equation}
where
\begin{equation}
    \label{A22}
    \begin{split}
    \Theta_{\bar{i}\bar{j}} (\rho, \epsilon) &= \mathring{\Theta}_{\bar{i}\bar{j}} (\rho\bmod{N_0}, \epsilon\bmod{N_1})
    \\
    \mathring{\Theta}_{\bar{i}\bar{j}} &= \Re\;[\textbf{DFT}(\widehat{S}^* (\widehat{V^{\bar{i}\bar{j}}})^*];
    \end{split}
\end{equation}
(\textbf{ii}) the component for ``scaling",
\begin{equation}
    \label{A23}
    \mathbf{b}_{\bar{r}\bar{s} 00} = \Theta_{\bar{r}\bar{s}} (0, 0)
    \end{equation}
where
\begin{equation}
    \label{A24}
    \begin{split}
    \Theta_{\bar{r}\bar{s}} (\rho, \epsilon) &= \mathring{\Theta}_{\bar{r}\bar{s}} (\rho\bmod{N_0}, \epsilon\bmod{N_1})
    \\
    \mathring{\Theta}_{\bar{r}\bar{s}} &= \Re\;[\textbf{DFT}(\widehat{S}^* (\widehat{U^{\bar{r}\bar{s}}})^*].
    \end{split}
\end{equation}

The lower block of $\mathbf{b}$ for ``background",
\begin{equation}
    \label{A25}
    \begin{split}
    \mathbf{b}_{\bar{p}\bar{q}} &= \Delta_{\bar{p}\bar{q}} (0, 0),
    \end{split}
\end{equation}
where
\begin{equation}
    \label{A26}
    \begin{split}
    \Delta_{\bar{p}\bar{q}} (\rho, \epsilon) &= \mathring{\Delta}_{\bar{p}\bar{q}} (\rho\bmod{N_0}, \epsilon\bmod{N_1})
    \\
    \mathring{\Delta}_{\bar{p}\bar{q}} &= N_0N_1 \Re\;[\widehat{S}^* \widehat{W^{\bar{p}\bar{q}}}]
    \end{split}
\end{equation}


\clearpage
\bibliography{JWST_SFFT}{}
\bibliographystyle{aasjournal}



\end{document}